\documentclass[10pt, conference, letterpaper]{IEEEtran}
\IEEEoverridecommandlockouts \IEEEaftertitletext{\vspace{-0.2in}}

\usepackage{amssymb}
\usepackage{amsmath}
\usepackage{mathtools}
\usepackage{latexsym}
\usepackage{graphicx}
\usepackage{subcaption}
\usepackage{mathtools}
\usepackage{latexsym}
\usepackage{algorithmic}
\usepackage{color}
\usepackage{multirow}
\usepackage{bm}
\usepackage{fancyhdr}
\usepackage[vlined, ruled, linesnumbered]{algorithm2e}

\newcommand{\kTO}       {{\bf to }}

\newcommand{\kETAL}    {{\em et al. }}

\newcommand{\mN}        {{\mathcal{N}}}

\newcommand{\ba}        {\textbf{a}}
\newcommand{\bs}        {\textbf{s}}

\newcommand{\bB}        {\textbf{B}}

\newcommand{\bG}        {\textbf{G}}
\newcommand{\bV}        {\textbf{V}}
\newcommand{\bE}        {\textbf{E}}
\newcommand{\bP}        {\textbf{P}}
\newcommand{\bp}        {\textbf{p}}
\newcommand{\ptotal}    {p_\text{total}}
\newcommand{\tbase}     {\text{base}}

\newcommand{\btheta}    {\bm{\theta}}

\DeclareMathOperator*{\argmax}{argmax}

\newenvironment{enum-algo}
{
    \begin{list}{Step \arabic{enumi}}
    {
    \usecounter{enumi}
    \setlength{\itemindent}{0.2em}
    \setlength{\labelwidth}{0.4in}
    \setlength{\leftmargin}{\labelwidth}
    }
} {
    \end{list}
}

\begin{document}

\title{Experience-driven Networking: A Deep Reinforcement Learning based Approach}
\author{Zhiyuan Xu, Jian Tang, Jingsong Meng, Weiyi Zhang, Yanzhi Wang, Chi Harold Liu and Dejun Yang
\thanks{Zhiyuan Xu, Jian Tang, Jingsong Meng and Yanzhi Wang
are with Department of Electrical Engineering and Computer Science, Syracuse University, Syracuse, NY 13244, USA.
Email: \{zxu105, jtang02, jmeng02, ywang393\}@syr.edu. Weiyi Zhang is with AT\&T Labs Research, Middletown, NJ 07748 USA.
Chi Harold Liu is with Beijing Institute of Technology, Beijing, China, 100081.
Dejun Yang is with Department of Electrical Engineering and Computer Science,
Colorado School of Mines, Golden, CO 80401, USA.
This research was supported in part by NSF grants 1704662, 1525920 and 1443966.
The information reported here does not reflect the position
or the policy of the federal government.
}}
\maketitle
\thispagestyle{fancy}
\begin{abstract}
%Traditional network resource allocation methods are either simple and straightforward (such as routing via minimum hop-count paths)
%or model-based, which assume network environment and user demands
%can be well modeled.  However,
%_{}
Modern communication networks have become very complicated and highly dynamic,
which makes them hard to model, predict and control.
In this paper, we develop a novel experience-driven approach that can
learn to well control a communication network from its own experience rather than
an accurate mathematical model, just as a human learns a new skill
(such as driving, swimming, etc). Specifically, we, for the first time,
propose to leverage emerging Deep Reinforcement Learning (DRL) for
enabling model-free control in communication networks; and present a novel
and highly effective DRL-based control framework, DRL-TE, for a fundamental networking problem:
Traffic Engineering (TE). The proposed framework maximizes a widely-used utility function
by jointly learning network environment and its dynamics, and making decisions under the
guidance of powerful Deep Neural Networks (DNNs).
We propose two new techniques, TE-aware exploration and actor-critic-based prioritized experience replay,
to optimize the general DRL framework particularly for TE.
To validate and evaluate the proposed framework, we implemented it in ns-3,
and tested it comprehensively with both representative and randomly generated network topologies.
Extensive packet-level simulation results show that 1) compared to several widely-used baseline methods,
DRL-TE significantly reduces end-to-end delay and consistently improves the network utility, while offering
better or comparable throughput; 2) DRL-TE is robust to network changes; and
3) DRL-TE consistently outperforms a state-of-the-art DRL method (for continuous control),
Deep Deterministic Policy Gradient (DDPG), which, however, does not offer satisfying performance.

\end{abstract}

%%%%%%%%%%%%%%%%%%%%%%%%%%%%%%%%%%%%%%%%%%%%%%%%%%%%%%%%%%%%%%%%%%%%%%%%%%%%%%%%%%%%%%%%%%%%%%%
%%%%%%%%%%%%%%%%%%%%%%%%%%%%%%%%%%%%%%  Section  %%%%%%%%%%%%%%%%%%%%%%%%%%%%%%%%%%%%%%%%%%%%%%
%%%%%%%%%%%%%%%%%%%%%%%%%%%%%%%%%%%%%%%%%%%%%%%%%%%%%%%%%%%%%%%%%%%%%%%%%%%%%%%%%%%%%%%%%%%%%%%
%==========================================================
\section{Introduction}
\label{Sec:Intro}
%===========================================================
Extensive research efforts have been made to develop algorithms
and protocols for communication networks to utilize their resources
efficiently and effectively. Traditional network resource allocation
methods are mostly model-based, which assume network environment and
user demand can be well modeled. However, communication networks have become
more complicated and highly dynamic, which makes them hard to model, predict
and control. Hence, we aim to develop a novel experience-driven model-free
approach that can learn to well control a communication network from its experience
rather than an accurate mathematical model, just as a human learns a skill (such as driving, swimming, etc).
We believe that some emerging networking technologies, such as Software Defined Networks (SDNs)~\cite{McKeown08},
can well support such an experience/data driven approach. For example, the Openflow controller
in an SDN can serve as the central control unit for collecting data, making decisions
and deploying solutions.

A fundamental networking problem is the Traffic Engineering (TE): given a set of network flows
with source and destination nodes, find a solution to forward the data traffic with the
objective of maximizing a utility function.
Simple and widely-used solutions include always routing traffic via shortest paths (e.g., Open Shortest Path First (OSPF)~\cite{OSPF});
or evenly distributing traffic via multiple available paths (e.g., Valiant Load Balancing (VLB)~\cite{Zhang-Shen10}). Obviously, neither of them
are optimal. Better solutions could be developed if there exist accurate and mathematically solvable
models for network environment, user demands and their dynamics.
Queueing theory has been employed to model communication networks and assist resource allocation~\cite{Li11,Paganini05,Palomar06,Xu11}.
However, it may not work well for those networking problems involving multi-hop routing and end-to-end performance (such as delay)
due to the following reasons: 1) In the queueing theory, many problems in a queueing network (rather than a single queue)
remain open problems, while a communication network with a mesh-like topology represents a fairly complicated multi-point to multi-point
queueing network where data packets from a queue may be distributed to multiple downstream queues, and a queue may
receive packets from multiple different upstream queues.
2) The queueing theory can only provide accurate estimations for queueing delay under a few strong assumptions
(e.g, tuple arrivals follow a Poisson distribution, etc), which, however, may not hold in a complex communication network.
Note that even if the packet arrival at every source node follows a Poisson distribution, packet arrivals at intermediate nodes may not.

In addition, Network Utility Maximization (NUM)~\cite{Low99} has been well studied,
which usually provides a resource allocation solution by formulating and
solving an optimization problem.
However, these methods may suffer from the following issues:
1) They usually assume that some key factors (such as user demands,
link usages, etc) are given as input, which, however, are hard to estimate or predict.
2) It is hard to directly minimize end-to-end delay by explicitly including it in the utility function
since given decision variables for resource allocation (such as TE), it is hard to express
the corresponding end-to-end delay in a closed form with them since an accurate mathematical model is needed to achieve this
(while queueing theory may not work here as described above).
3) Network dynamics have not been well addressed by these works.
Most of them claimed to provide a ``good'' resource allocation solution, which
is optimal or close-to-optimal but only for a snapshot of the network.
However, most communication networks are highly time-varying. How resource allocation should be adjusted
or re-computed to accommodate such dynamics has not been well addressed by these NUM methods.

Recent breakthrough of \emph{Deep Reinforcement Learning (DRL)}~\cite{Mnih15} provides a promising technique
for enabling effective experience-driven model-free control.
\emph{DRL} (originally developed by DeepMind)
enables computers to learn to play games, including Atari 2600 video games and one of the most complicated games, Go (AlphaGo~\cite{Silver16}),
and beat the best human players.
Even though DRL has made tremendous successes on game-playing that usually has a limited action space (e.g., moving up/down/left/right),
it has not yet been investigated how DRL can be leveraged for resource allocation problems (such as TE)
in complex communication networks, which usually have sophisticated states and huge or continuous action spaces.

%DRL consists of an offline Deep Neural Networks (DNN) construction phase, which correlates the value function with
%corresponding states and actions, and an online deep Q-learning phase for action selection, system control,
%and DNN updating.
%
We believe DRL is especially promising for control in communication networks because:
1) It has advantages over other dynamic system control techniques such as model-based predictive control
in that the former is model-free and does not rely on accurate and mathematically solvable system models (such as queueing models),
thereby enhancing its applicability in complex networks with random and unpredictable behaviors.
2) It is able to deal with highly dynamic time-variant environments such as time-varying system states and user demands.
3) It is capable of handling a sophisticated state space (such as AlphaGo~\cite{Silver16}),
which is more advantageous over traditional Reinforcement Learning (RL)~\cite{Sutton98}.
%
%
%4) A DRL-based control framework usually has a low online computational complexity.
%4) Model-free control needs to be guided by big data. DSDPSs are becoming larger and more complicated,
%generating a large amount of statistics data (such as workload, traffic load, resource usages, etc) every second.
%Instead of treating these data as an unwanted burden, DRL, can be leverage them to train DNNs that
%are able to make the best action selections at runtime.
%
However, direct application of the basic DRL technique, such as Deep Q-Network (DQN) based DRL (proposed in the pioneering work~\cite{Mnih15}),
does not work for the TE problem since it is a continuous control problem (See Section~\ref{Sec:DRL});
while DQN-based DRL is only capable of handling control problems with a limited action space. Although DRL methods have been
proposed for continuous control very recently~\cite{Gu16, Lillicrap16}, we show a state-of-the-art method, Deep Deterministic Policy Gradient (DDPG)~\cite{Lillicrap16},
does not work well for our TE problem.
%
%Moreover, the existing DRL methods~\cite{Mnih15} usually need to collect big data (e.g., lots of images for the game-playing
%applications) for learning, which are additional overhead and burden for an online system.
%Our goal is to develop a method that only needs to collect very limited statistics data during runtime.

In this paper, we develop a novel and highly effective DRL-based model-free control framework for TE in
a communication network to jointly learn network dynamics and making decisions under the guidance of
powerful Deep Neural Networks (DNNs).
We summarize our contributions in the following:

\begin{itemize}

\item We are the first to present a highly effective and practical DRL-based experience-driven control framework, DRL-TE, for TE.

\item We discuss and show that direct application of a state-of-the-art DRL solution for continuous control, namely Deep Deterministic Policy Gradient (DDPG)~\cite{Lillicrap16}, does not work well for the TE problem.

\item We propose two new techniques, TE-aware exploration and actor-critic-based prioritized experience replay
to optimize the general DRL framework particularly for TE.

\item We show via extensive packet-level simulation using ns-3~\cite{ns3} with both representative and random network topologies that
DRL-TE significantly outperforms several widely-used baseline methods.

\end{itemize}

\emph{To the best of our knowledge, we are the first to leverage the emerging DRL for enabling model-free control in communication networks.
We aim to promote a simple and practical experience-driven approach based on DRL, which, we believe,
can be easily extended to solve many other resource allocation problems in communication networks.}
%
%The rest of the paper is organized as follows:
%%
%We give a brief background introduction to DRL in Section~\ref{Sec:Background}.
%We describe the TE problem in Section~\ref{Sec:Problem}.
%We then present the proposed framework in Section~\ref{Sec:DRL}.
%%
%Simulation settings are described, and simulation results are presented and
%analyzed in Section~\ref{Sec:Eval}.
%%
%We discuss related work in Section~\ref{Sec:Related} and conclude the
%paper in Section~\ref{Sec:Conclusions}.

%%%%%%%%%%%%%%%%%%%%%%%%%%%%%%%%%%%%%%%%%%%%%%%%%%%%%%%%%%%%%%%%%%%%%%%%%%%%%%%%%%%%%%%%%%%%%%%
%%%%%%%%%%%%%%%%%%%%%%%%%%%%%%%%%%%%%%  Section  %%%%%%%%%%%%%%%%%%%%%%%%%%%%%%%%%%%%%%%%%%%%%%
%%%%%%%%%%%%%%%%%%%%%%%%%%%%%%%%%%%%%%%%%%%%%%%%%%%%%%%%%%%%%%%%%%%%%%%%%%%%%%%%%%%%%%%%%%%%%%%
%==========================================================
\section{Deep Reinforcement Learning (DRL)}
\label{Sec:Background}
%===========================================================
We provide necessary background about DRL in this section.
We consider a standard RL setup consisting of an agent interacting with an environment
in discrete decision epochs. At each decision epoch $t$, the agent observes state $\bs_t$, takes
an action $\ba_t$ and receives a reward $r_t$. The objective is to find
a policy $\pi(\bs)$ mapping a state to an action (deterministic) or a probability distribution
over actions (stochastic) with the objective of maximizing the discounted cumulative reward
$R_0 = \sum_{t=0}^{T}\gamma^t r(\bs_t,\ba_t)$, where $r(\cdot)$ is the reward function and
$\gamma \in [0,1]$ is the discount factor.

In the seminal work~\cite{Mnih15}, DeepMind introduced DRL, which extends the well-known Q-learning
to enable end-to-end system control based on high-dimensional sensory inputs (such as raw images).
The training phase adopts a DNN called Deep Q-Network (DQN) to derive the correlation
between each state-action pair $(\bs_t,\ba_t)$ of the system
under control and its value function $Q(\bs_t,\ba_t)$, which is the expected discounted cumulative reward.
If the system in state $\bs_t$ and follows action $\ba_t$ at decision epoch $t$ (and a certain policy $\pi$ thereafter):
\begin{equation}
Q(\bs_t,\ba_t)=\mathbb{E}\Big[R_t|\bs_t,\ba_t\Big],
\end{equation}
where $R_t = \sum_{k=t}^{T}\gamma^k r(\bs_t,\ba_t)$.
A commonly-used off-policy algorithm takes the greedy policy: $\pi(\bs_t) =  \argmax_{\ba_t} Q(\bs_t,\ba_t)$.
The DQN can be trained by minimizing the loss:
\begin{equation}
L(\btheta^{Q})=\mathbb{E}\Big[y_t - Q(\bs_t,\ba_t|\btheta^{Q}) \Big],
\label{Eqn:Loss}
\end{equation}
where $\btheta^{Q}$ is the weight vector of the DQN and $y_t$ is the target value, which can be estimated by:
\begin{equation}
y_t = r(\bs_t,\ba_t) + \gamma Q(\bs_{t+1},\pi(\bs_{t+1}|\btheta^{\pi})|\btheta^{Q}).
\label{Eqn:Target}
\end{equation}

It is not new to use a neural network (or even DNN) as the function approximator in RL. But
a non-linear function approximator (such as neural network) is known to be unstable or even to diverge.
Two effective techniques were introduced in~\cite{Mnih15} to improve stability: experience relay and target network.
Unlike traditional RL, a DRL agent updates the DNN with a mini-batch from an experience replay buffer~\cite{Mnih15}, which
stores state transition samples collected during learning.
Compared to using only immediately collected samples (such as original Q-learning), randomly sampling from the experience replay
buffer allows the DRL agent to break the correlation between sequentially generated samples, and learn from a more independently and identically distributed
past experiences, which is required by most of training algorithms, such as Stochastic Gradient Descent (SGD).
So experience replay can smooth out learning and avoid oscillations or divergence.
In addition, a DRL agent uses a separate target network (which has the same structure as the DQN) to estimate target values $<y_t>$
for training the DQN, whose parameters, however, are slowly updated with the DQN weights
every $C>1$ epochs and are held fixed between individual updates.

The traffic engineering problem (described next) is a continuous control problem. Unfortunately, the DQN-based DRL
only works for control problems with a low-dimensional discrete action space. It cannot be easily
applied to continuous control since it needs to find the action that maximizes the
action-value function, which, however, requires an iterative process
to solve a non-trivial non-linear optimization problem at each epoch.
A straightforward solution to adapting DQN-based approach to continuous
cases is to simply discretize the action space, which, however, may likely leads to
a huge number of actions, which are very hard to deal with too.

Continuous control has often been tackled by the actor-critic approach~\cite{Konda00}, which usually
employs the policy gradient method to search for the optimal policy.
The traditional actor-critic approach can also be extended to embrace DNN (such as DQN)
to guide decision making~\cite{Lillicrap16}.
For example, a recent work~\cite{Lillicrap16} from DeepMind introduced an actor-critic method, called
Deep Deterministic Policy Gradient (DDPG), for continuous control. The basic idea is to maintain a
parameterized actor function $\pi(\bs_t|\btheta^{\pi})$ and a parameterized critic function $Q(\bs_t, \ba_t|\btheta^{Q})$.
The critic function can be implemented using the above DQN, which returns $Q$ value
for a given state-action pair.
The actor function can also be implemented using a DNN, which specifies the current policy by mapping a state to
a specific action.  According to~\cite{Silver14}, the actor network can be updated by applying the chain
rule to the expected cumulative reward $J$ with respect to the actor parameters $\btheta^{\pi}$:
\begin{equation}
\begin{split}
\nabla_{\btheta^{\pi}} J & \approx \mathbb{E}\Big[ \nabla_{\btheta^{\pi}} Q (\bs, \ba|\btheta^{Q}) |_{\bs=\bs_t, \ba = \pi(\bs_t|\btheta^{\pi})} \Big]\\
& = \mathbb{E}\Big[ \nabla_{\ba} Q (\bs, \ba|\btheta^{Q}) |_{\bs=\bs_t, \ba = \pi(\bs_t)} \cdot \nabla_{\btheta^{\pi}} \pi (\bs|\btheta^{\pi}) |_{\bs=\bs_t} \Big].
\label{Eqn:DPG}
\end{split}
\end{equation}
Note that the experience replay and target network introduced above
can also be integrated to this approach to ensure stability.

%%%%%%%%%%%%%%%%%%%%%%%%%%%%%%%%%%%%%%%%%%%%%%%%%%%%%%%%%%%%%%%%%%%%%%%%%%%%%%%%%%%%%%%%%%%%%%%
%%%%%%%%%%%%%%%%%%%%%%%%%%%%%%%%%%%%%%  Section  %%%%%%%%%%%%%%%%%%%%%%%%%%%%%%%%%%%%%%%%%%%%%%
%%%%%%%%%%%%%%%%%%%%%%%%%%%%%%%%%%%%%%%%%%%%%%%%%%%%%%%%%%%%%%%%%%%%%%%%%%%%%%%%%%%%%%%%%%%%%%%
%==========================================================
\section{Problem Statement}
\label{Sec:Problem}
%===========================================================
We describe the Traffic Engineering (TE) problem in this section. First,
we summarize the major notations below for quick reference.

\begin{table}[!ht]
\caption{Notation Definition}
\begin{center}
\begin{tabular}{| c | c |}
\hline
Variable & Definition \\
\hline
$K$ & The number of communication sessions \\
\hline
$\bP_k$  & The set of candidate paths of session $k$ \\
\hline
$\bE$  & The set of links of the network\\
\hline
$B_k$ & Traffic demand of session $k$ \\
\hline
$C_e$ &Capacity of link $e$ \\
\hline
$f_{k,j}$ & The amount of traffic of \\& the $j$th path of session $k$ \\
\hline
$w_{k,j}$ & Split ratio for the $j$th path of session $k$ \\
\hline
$x_k$, $z_k$ & Throughput and delay of session $k$ \\
\hline
$\bs$, $\ba$, $r$ & State, action and reward\\
\hline
$p_i, P(i)$ & Priority and probability (being selected)\\ & of transition sample $i$\\
\hline
$\btheta{^\pi}, \btheta^Q$ & Weights of actor and critic networks $\pi(\cdot)$ and $Q(\cdot)$\\
\hline
\end{tabular}
\end{center}
\end{table}
%
%Network model
We consider a general communication network with $K$ end-to-end communication sessions.
We use a directed graph $\bG(\bV,\bE)$ to model the network, where each vertex corresponds to
a node (router or switch) and each edge corresponds to a directed
communication link connecting a pair of nodes.
Each communication session $k$ has a source node $s_k$, destination $d_k$ and a set of candidate paths $\bP_k$  (connecting $s_k$ with $d_k$)
that can carry its traffic load. As mentioned above, we aim to study a TE problem
seeking a rate allocation solution, which specifies the amount of traffic load $f_{k,j}$ going through the $j$th
path of $\bP_k$. Note that once we have such a solution, then when a packet of session $k$ arrives at
$s_k$, path $j$ is chosen to transmit the packet with a probability of $w_{k,j}$,
where $w_{k,j}=f_{k,j} / (\sum_{j=1}^{|P_k|}f_{k, j})$, which is known as the split ratio.
%
%We calculate the total throughput $x_k$ of communication session $k$,
%defined as the total number of bytes received divided by the statistical time.
%We also calculate the average packet-level end-to-end delay $z_k$ within a statistical time.
%Due to the naturally dynamical property of networks
%(e.g, dynamical packet arrival rate, queuing delay at intermediate nodes, different allocation rate, etc),
%we redo the calculation every statistical time (in our framework, is one decision epoch).

%Objective function
The $\alpha$-fairness~\cite{Srikant12,Winstein13} model has been widely used for NUM.
According to this model, the utility of a communication session with a steady-state throughput of $x$
is $U_{\alpha}(x)=(\frac{x^{1-\alpha}}{1-\alpha}$). Particularly, as $\alpha \to 1$, in the limit $U_1{(x)}$ becomes $\log x$~\cite{Winstein13}.
For $\alpha > 0$, $U_{\alpha}(x)$ is monotonically increasing with $x$.
%
%an allocation that maximizes the total score prefers to divide the throughput of a bottleneck link equally between communication sessions.
%
The objective of the TE problem is usually set to maximizing the total utility of
all the communication sessions, i.e., $\sum_{k=1}^K {U_{\alpha}(x)}$.
$\alpha$ can be used to tradeoff fairness and efficiency.
%
%If $\alpha=0$, the objective is simply to maximize total throughput without consideration for fairness.
%In contrast, if $\alpha=2$, the obje is set to minimum delay fairness, since the the scores goes as the negative inverse of the throughput.
%
%As $\alpha \to \infty$, the objective is to achieve max-min fairness since
%all that matters is the minimum allocation~\cite{Srikant12}.
%
If $\alpha=1$, the objective is to achieve the proportional fairness, which is widely used for resource allocation.
%
%it will cut one user's allocation in half as long as another user's can be more than doubled.

In order to address both throughput and delay, similar as in~\cite{Winstein13},
we define a utility function $U(\cdot)$ for session $k$:
\begin{equation}
U(x_k, z_k) = U_{\alpha_1}(x_k) - \sigma \cdot U_{\alpha_2}(z_k),
\end{equation}
%
%$\alpha$ and $\beta$ are set to tradeoff fairness and efficiency tradeoff between communication session throughput and delay, respectively,
where $x_k$ and $z_k$ are the end-to-end throughput and delay of session $k$ respectively; and
$\sigma$ expresses the relative importance of delay vs. throughput.
%
%Note that we set $\alpha=1$ in our case, which means to achieve a balance between fairness and efficiency.
%
Similarly, the objective of the TE problem is to maximize the total utility
of all the communication sessions in the network, i.e., $\sum_{k=1}^K {U(x_k,z_k)}$.
%
%we choose $\sum_kU(x_k, z_k)$ as our reward function to guide the DRL agent to make decisions.
%It's worth mentioning that the purpose of the utility function is to provide guidance for DRL agent and quantitatively evaluate different control methods.
%Maybe in practical scenario, we have a different objective,
%for example, an online video chat cares more about delay rather than throughput.
%But, as we will show later, our DRL-based framework can flexibly deal with different utility functions.

Note that we aim to consider a general communication network and show how DRL can
enable experience-driven networking rather than targeting at a specific physical
network (such as SDN, multihop wireless network) or a specific scenario (such as WAN, MAN, LAN, etc).
So we try to make system model and problem statement as general as possible. However,
the proposed control framework (Section~\ref{Sec:DRL}) is so flexible that it can be easily extended
for a specific network or scenario with additional constraints.
%
%which will be discussed further in the next section.

%%%%%%%%%%%%%%%%%%%%%%%%%%%%%%%%%%%%%%%%%%%%%%%%%%%%%%%%%%%%%%%%%%%%%%%%%%%%%%%%%%%%%%%%%%%%%%%
%%%%%%%%%%%%%%%%%%%%%%%%%%%%%%%%%%%%%%  Section  %%%%%%%%%%%%%%%%%%%%%%%%%%%%%%%%%%%%%%%%%%%%%%
%%%%%%%%%%%%%%%%%%%%%%%%%%%%%%%%%%%%%%%%%%%%%%%%%%%%%%%%%%%%%%%%%%%%%%%%%%%%%%%%%%%%%%%%%%%%%%%
%==========================================================
\section{Proposed DRL-based Control Framework}
\label{Sec:DRL}
%===========================================================
In this section, we present the proposed DRL-based control framework, DRL-TE, for
the TE problem described above.

In order to utilize the DRL techniques (no matter which method/model to use), we first
need to design the state space, action space and reward function.
%
%We have introduced the general framework of DRL-based control in section~\ref{Sec:DRL}.
%Before we can apply DRL to solve the above traffic engineering problem,
%we first need to design the state, action and reward.
%
\begin{itemize}
\item \emph{State Space}: The state consists of two components: throughput and delay of each communication session.
Formally, the state vector $\bs = [(x_1,z_1), \cdots, (x_k, z_k), \cdots, (x_K, z_K)]$.
%
%In order to get a better representation of state, we apply normalization first to each dimension.

\item \emph{Action Space}: An action is defined as the solution to the TE problem, i.e., the set of split ratios for the communication sessions.
Formally, the action vector $\ba = [w_{1,1}, \cdots, w_{kj}, \cdots, w_{K,|P_k|}]$, where $\sum_{j=1}^{|P_k|}w_{k,j}=1$.

\item \emph{Reward}: The reward is the objective of the TE problem, which is the total utility of all the communication sessions.
Formally, $r=\sum_{k=1}^K {U(x_k, z_k)}$.
\end{itemize}
%
%This constrain ensures the sum of split ratios in the same communication session is 1.
%
%Straightforwardly, we take the sum of utility function value for each communication session as immediate reward ,
%we choose $\alpha=\beta=1$ which corresponds to proportional throughput and delay fairness, and try different $\mu$ value to see how our DRL agent can adjust its solution according to different objective.
%Overall, the immediate reward $r$ is,
%
%\begin{equation}
%r = \sum_k(\log x_k - \mu \cdot \log y_k)
%\end{equation}

Note that the design of state space, action space and reward is critical to the success of a DRL method.
Our design well captures network states and the key components of the TE problem without including
useless/redudant information.
The core of the proposed control framework is an agent, which runs a DRL algorithm (Algorithm~\ref{Alg:drl}) to find the best action at
each decision epoch, takes the action to the network (e.g., through a network controller)
observes the network state, and collects a transition sample.
%
%We observed that based on our design, the proposed DRL based framework can well model the correlation between a allocation solution (state) and overall utility value (reward) after training.
%There is no much help for adding more fine-grid information, such as throughput or delay for each path in a session.

The TE problem is obviously a continuous control problem. As explained above, the DQN-based DRL proposed
in the well-known work~\cite{Mnih15} does not work here; so we choose the state-of-the-art
DRL-based solution for continuous control, DDPG~\cite{Lillicrap16}, as the starting point for our design,
whose basic idea has been introduced in Section~\ref{Sec:Background}.
Even though DDPG has been demonstrated to work well on quite a few continuous control tasks~\cite{Lillicrap16}, our experimental results,
however, show that direct application of DDPG to the TE problem does not lead to satisfying performance (Section~\ref{Sec:Eval}).
We suspect this is due to the following two reasons:
1) The DDPG framework in~\cite{Lillicrap16} does not clearly specify how to explore.
A simple random noise based method or the exploration methods proposed for
physical control problems (mentioned in~\cite{Lillicrap16}) do not work well for the TE problem here.
2) DDPG utilizes a simple uniform sampling method for experience replay, which
ignores the significance of transition samples in the replay buffer.
To address these two issues, we propose two new techniques to optimize DDPG particularly for TE,
including TE-aware exploration which leverages a good TE solution as the baseline during
exploration; and actor-critic-based prioritized experience replay which can employs a new method
for specifying significance of samples with careful consideration for both the actor and critic networks.
Exploration is an essential and important process for training a DRL agent because
an inexperienced agent needs to see sufficient transition samples to
gain experience and eventually learn a good (hopefully optimal) policy.
For continuous control problems, exploration is quite challenging because
there are infinite number of actions that can be chosen in each decision epoch and the commonly-used
$\epsilon$-greedy method~\cite{Mnih15} only works for tasks with a limited discrete action space, which
obviously does not work here. DDPG generates an action for exploration
by adding a random noise to the action returned by the current actor network.

For exploration, we propose a new randomized algorithm that guides the exploration process
with a base TE solution.
Specifically, with $\epsilon$ probability, the DRL agent derives action as $\ba_{\tbase} + \epsilon \cdot \mN$;
and with $(1-\epsilon)$ probability, it derives action as $\ba+\epsilon \cdot \mN$;
where $\ba_{\tbase}$ is a base TE solution, $\ba$ is the output of actor network $\pi(\cdot)$ and $\epsilon$ is an adjustable parameter.
$\epsilon$ can tradeoff exploration and exploitation by determining the probability of adding a random noise to the action
rather than taking the derived action from the actor network.
$\epsilon$ decays with decision epoch $t$, which means with more learning,
more derived (rather than random) actions  will be taken.
The parameter $\mN$ is a uniformly distributed random noise.
%
%each element of which was set to a random number in
%$[0,0.3]$ in our implementation.

The proposed control framework is not restricted to any specific
base TE solution for $\ba_{\tbase}$, which can be obtained in many different ways.
For example, a simple solution is to use the shortest path to deliver all the packets for
each communication session, which is not optimal in most cases but is good enough to sever
as a baseline for exploration. Another solution is to
evenly distribute traffic load of each communication session to all candidate paths.
NUM-based methods can also be used to find base solutions. For example, we can obtain a TE solution by solving the following
mathematical programming:
%%
%The exploration is essential to DRL agent during the training process.
%Since we know some priori knowledge about network configuration, we can formulate a linear-programing (LP) problem, and derive a base solution easilty.
%Note that we only need to provide a basic one to DRL, rather than an optimal one.
%Our actual objective is to maximize the overall utility function, and we need to make a tradeoff between throughput and delay.
%However, it's quite difficult to formulate the real network,
%we need to include user demand, link capacity, queuing delay and so on.
%Thus, we only consider a max-flow problem, which only tries to maximize the overall throughput in an fair way.
%This is a light way for design and resolve, and the DRL can take the advantage of base solution as well. The variables used in LP problem is summarized in table~\ref{}.

\noindent NUM-TE:
\begin{subequations}
\begin{align}
\underset{<x_k, f_{k,j}>}{\max} & \sum_k{U_{\alpha}(x_k)} \\
\nonumber \text{subject to:}\\
\sum_{k=1}^K {\sum_{\bp_j \in \bP_k: e \in \bp}{f_{k,j}}} \leq C_e, &~\forall~e \in \bE;
\label{Eqn:C}\\
x_k \leq B_k, &~k \in \{1,\cdots,K\};
\label{Eqn:B}\\
\sum_{j=1}^{|\bP_k|}{f_{k,j}} = x_k, &~k \in \{1,\cdots,K\}.
\label{Eqn:fx}
\end{align}
\end{subequations}
In this formulation, the objective is to maximize the total utility in terms of throughput. Note that it is hard to
include the end-to-end delay term in the utility function since there does not exists a mathematical model that can accurately establish a
connection between end-to-end delay and the other decision variables $<x_k, f_{k,j}>$. This is why end-to-end delay has not
been well addressed by most existing works on NUM.
Constraints~(\ref{Eqn:C}) ensure the aggregated traffic load on each link does not exceed its capacity $C_e$, where $\bp_j$ is the $j$th path
in $\bP_k$.
Constraints~(\ref{Eqn:B}) ensure the total throughput of each session $k$ does not exceed its demand $B_k$, which can be estimated.
Constraints~(\ref{Eqn:fx}) establish the connections between two set of decision variables $<x_k>$ and $<f_{k,j}>$.
%
%We use variable $x_k$ to express the aggregated data rate for session $k$ (10d).
%
%
If $\alpha=1$, $U_{\alpha}(x_k)=\log{x_k}$, then this problem becomes a convex programming problem, which can be efficiently
solved by the Gurobi Optimizer~\cite{Gurobi} that were used in our implementation.

%Although this DDPG-based method can provide solutions to our continuous control problem
%and does achieve model-free control for resource allocation,
%it faces the following issues.
%
DDPG simply uniformly samples transition data from the experience replay.
It has been shown by~\cite{Schaul15} that an DRL agent can learn more effectively from some transitions than others.
%
%Some transition samples may not be immediately useful to the agent, but might become important later.
%
A method called prioritized experience replay has also been introduced in~\cite{Schaul15},
which has been shown to lead to better performance on game-playing tasks when being
combined with DQN.
It assigns a priority for each transition sample. Based on this priority, transition data in
the replay buffer are sampled in each epoch.
However, this method was proposed only for DQN-based DRL and has never been used
with the actor-critic method for continuous control.
We extend this method to enable prioritized experience replay under the actor-critic framework.
%
%In our traffic engineering problem, we can have some priori knowledge about network, such as link capacity, paths, etc..
%The DRL agent can explore around a basic solution based on those knowledge, this will accelerate the speed of finding a good solution.
%
Specifically, since an actor-critic method uses two networks (actor and critic) to guide decision making,
the priority should consist of two parts.
The first part is the Temporal-Difference (TD) error, which corresponds to training of the critic network:
%
%The TD error $\delta$ is defined as
\begin{equation}
\delta = y - Q(\bs, \ba),
\end{equation}
where $y$ is the target value for training the critic network, which is defined in Equation~(\ref{Eqn:Target}).
Note that to help understand the basic idea better, we omit the subscripts/superscripts here for clean presentation;
the exact forms of these equations can be found at the formal algorithm presentation.
The actor and critic network are jointly trained by transition samples in the replay buffer.
The second part is related to training of the actor network, i.e.,
the $Q$ gradient $\nabla_{\ba} Q = \nabla_{\ba} Q (\bs, \ba) |_{\bs=\bs_i, \ba = \pi (\bs_i)}$  (Equation~(\ref{Eqn:DPG})).
Combining them together, the priority of a transition sample is given as:
\begin{equation}
p = \varphi \cdot (|\delta|+\xi) + (1-\varphi) \cdot \overline{|\nabla_{\ba} Q|},
\label{Eqn:Priority}
\end{equation}
where $\varphi$ is a parameter controlling the relative importance of TD error vs. $Q$ gradient.
$\overline{|\nabla_{\ba} Q|}$ is the average of absolute values of the $Q$ gradient (which is a vector).
A small positive constant $\xi$ is used to prevent the edge-cases of transitions not being revisited once their error is zero.
The probability of sampling transition $i$ is:
\begin{equation}
P(i)=\frac{p_i^{\beta_0}}{\sum_j^{|\bB|} p_j^{\beta_0}},
\label{Eqn:Prob}
\end{equation}
where the exponent $\beta_0$ determines how much prioritization is used; if
$\beta_0=0$, then it becomes uniform sampling.

\begin{algorithm}[!ht]
\caption{DRL-TE}
\label{Alg:drl}
\begin{algorithmic}[1]
% \item \emph{Input}: training steps $T$, exponents $\beta_0$ and $\beta_1$, reward discount $\gamma$, priority weight $\varphi$, step size $\eta^Q$ and $\eta^{\pi}$, target update rate $\tau$.
%
\STATE Randomly initialize critic network $Q(\cdot)$ and actor network $\pi(\cdot)$ with weights $\bm{\theta^{Q}}$ and $\bm{\theta^{\pi}}$ respectively;
\STATE Initialize target networks $Q'(\cdot)$ and $\pi'(\cdot)$ with weights $\bm{\theta^{Q'}} := \bm{\theta^{Q}}$, $\bm{\theta^{\pi'}} := \bm{\theta^{\pi}}$;
\STATE Initialize prioritized replay buffer $\bB$ and $p_1:=1$; \\
% /**Offline Training**/
% \STATE Load the historical transition samples into $\bB$, train the actor and critic network offline; \\
/**Online Learning**/
%\STATE Initialize a random process $\mathcal{R}$ for exploration;
\STATE Receive the initial observed state $\bs_1$;\\
/**Decision Epoch**/
\FOR{t = 1 \kTO $T$}
  %\STATE \Red{Derive action $\ba_t$ from the actor network $\pi(\bs_t)$;}
  \STATE Apply the TE-aware exploration method to obtain $\ba_t$;
  \STATE Execute action $\ba_t$ and observe the reward $r_t$;
  \STATE Store transition sample $(\bs_t, \ba_t, r_t, \bs_{t+1})$ into $\bB$ with maximal priority $p_t=\text{max}_{j<t}~p_j$;
  \STATE /**Prioritized Transition Sampling**/
  \FOR{i = 1 \kTO $N$}
  \STATE Sample a transition $(\bs_i, \ba_i, r_i, \bs_{i+1})$ from $\bB$ where $i\sim P(i) := p_i^{\beta_0} / \sum_j p_j^{\beta_0}$;
  \STATE Compute important-sampling weight: $\omega_i := (|\bB| \cdot P(i))^{-\beta_1}/\text{max}_j \omega_j$;
  \STATE Compute target value for critic network: $Q(\cdot)$ $y_i := r_i + \gamma \cdot Q' (\bs_{i+1},\pi' (\bs_{i+1}))$;
  \STATE Compute TD-error: $\delta_i := y_i - Q(\bs_i, \ba_i)$;
  \STATE Compute gradient: $\nabla_{\bm{\btheta^{\pi}}} J_i :=$ $\nabla_{\ba} Q (\bs, \ba) |_{\bs=\bs_i, \ba = \pi (\bs_i)} \cdot \nabla_{\bm{\btheta^{\pi}}} \pi (\bs) |_{\bs=\bs_i}$;
  \STATE Update the transition priority: $p_i := \varphi \cdot (|\delta_i|+\xi) + (1-\varphi) \cdot \overline{|\nabla_{\ba} Q|}$;
  \STATE Accumulate weight-change for critic network: $Q(\cdot)$ $\Delta_{\btheta^Q} := \Delta_{\btheta^Q} + \omega_i \cdot \delta_i \cdot \nabla_{\btheta^Q}Q(\bs_i,\ba_i)$;
  \STATE Accumulate weight-change for actor network: $\pi(\cdot)$ $\Delta_{\btheta^{\pi}} := \Delta_{\btheta^{\pi}} + \omega_i \cdot \nabla_{\bm{\btheta^{\pi}}} J_i $;
  \ENDFOR
  % \STATE Update the critic network $Q(\cdot)$ by minimizing the loss: \\
  % $L(\bm{\theta^Q}) = \frac{1}{N} \sum\limits_{i=1}^{N}(w_i \cdot \delta_i)^2$;
  \STATE /**Network Updating**/
  \STATE Update the weights of critic network: $Q(\cdot)$ $\btheta^Q := \btheta^Q + \eta^Q \cdot \Delta_{\btheta^Q}$,~reset $\Delta_{\btheta^Q} := 0$;
  \STATE Update the weights of actor network: $\pi(\cdot)$ $\btheta^{\pi} := \btheta^{\pi} + \eta^{\pi} \cdot \Delta_{\btheta^{\pi}}$,~reset $\Delta_{\btheta^{\pi}} := 0$;
  % \STATE Update the weights $\bm{\theta^{\pi}}$ of actor network $f(\cdot)$ using the sampled gradient: \\
  % $\nabla_{\bm{\theta^{\pi}}} f \approx$ $\frac{1}{N} \sum\limits_{i=1}^N \nabla_{\ba} Q (\bs, \ba) |_{\ba = f (\bs_i)} \cdot \nabla_{\bm{\theta^{\pi}}} f (\bs) |_{\bs_i}$; \\
  \STATE Update the weights of the corresponding target networks: \\
  $\btheta^{Q'} := \tau \btheta^{Q} + (1 - \tau) \btheta^{Q'}$; \\
  $\btheta^{\pi'} := \tau \btheta^{\pi} + (1 - \tau) \btheta^{\pi'}$; \\
\ENDFOR
\end{algorithmic}
\end{algorithm}

We formally present the proposed DRL-based control framework for TE, DRL-TE, as Algorithm~\ref{Alg:drl}.
First the algorithm randomly initializes all the weights $\btheta^\pi$ of actor network $\pi(\cdot)$; and $\btheta^Q$ of the critic networks $Q(\cdot)$(line 1).
%
%As shown in \cite{}, if we directly apply the actor and critic network to generate the target value $y_i$ (line 15),
%it may suffer from unstable and divergence problems.
%
As mentioned above, we employ target networks $\pi'(\cdot)$ and $Q'(\cdot)$ to improve learning stability.
The target networks are clones of the original actor or critic networks,
whose weights $\btheta^{\pi'}$ and $\btheta^{Q'}$ are initialized in the same way as their original networks (line 2)
but are slowly following updated (line 23). The update rate is controlled by a parameter $\tau$.
In each decision epoch, %DRL agent derives an action from actor network according to its current state (line 7),
the algorithm applies the TE-aware exploration method to obtain an action first (line 6), which is explained above.
%In our implementation, we set $\tau=0.01$.

We use a prioritized replay buffer for storing transition samples.
%
%Instead of training network using the transition sample immediately collected
%at each decision epoch $t$ (line 7-10),
%
We first store the sample into the replay buffer with maximal priority (line 8),
and then sample a mini-batch of transition samples from $\bB$ (lines 10-19) to train the actor and critic networks.
The priority of transition is then updated using the method described right above (lines 16-18).
Note that for every transition sample $(\bs_i, \ba_i, r_i, \bs_{i+1})$ in the mini-batch,
we first obtain its important-sampling weight $\omega$ (line 12), which is used to correct the bias introduced by prioritized replay~\cite{Schaul15}.
The weight is integrated into the critic network updating in the form of $\omega \cdot \delta$ (rather than $\delta$ only) (line 17).
Priorities ensure high-error transitions are seen more frequently. Those large steps (with large priority) can be very disruptive
because of large updating values. As suggested by~\cite{Schaul15}, we import annealing weight $\beta_1$ to correct this bias,
by linearly annealing it from its initial value to 1 (line 12). For learning stability,
we always normalize $\omega$ by $1/\text{max}_j \omega_j$, so they only scale the weight update downward.
We obtain the action for the next state from target actor network $\pi'(\bs_{i+1})$,
and the target value $y_i$ (line 13) for training the critic network;
%
%The discount factor $\gamma=0.99$ in our implementation;
%
In addition, we compute the policy gradient by the chain rule, as described in Equation~(\ref{Eqn:DPG}) (line 15).
%
%The actor and critic network $Q(\cdot)$ are trained by the prioritized mini-batch samples from $B$ as mentioned above.
The weight-changes are accumulated (lines 17-18) and used to update the actor and critic networks (lines 21-22).

There are quite a few hyper-parameters in the proposed control framework.
To maximize its performance, we conducted a comprehensive empirical study
to find the best settings for them and the best structures of the actor and critical networks.
In our design and implementation, we used a 2-layer fully-connected feedforward neural network to serve as the actor network,
which includes $64$ and $32$ neurons in the first and second layer respectively and utilized
the Leaky Rectifier~\cite{Goodfellow16} for activation.
%
%\Red{we aggregate $K$ neurons as the final output layer, and in each neuron, it has $|P_k|$ outputs and uses softmax as activation function to ensure the sum of output value for communication session $k$ equals one.}
%
In the final output layer, we, however, employed the softmax~\cite{Goodfellow16} as activation function
to ensure the sum of output values equals one.
%
%Note that in our traffic engineering problem, we need to decide the split ratio for each path on a communication session,
%and we have constrain that the sum of split ratios on the same session is 1.
%Thus, we aggregate $K$ neurons as the output layer, and for each neuron, it has $P_k$ outputs and uses softmax as activation function,
%where $K$ is the number of communication sessions, and $P_k$ is the number of paths on communication session $k$.
%In this way, we can ensure that the output action $\ba$ meets our sum constrains.
%
For the critic network, we also used a 2-layer fully-connected feedforward neural network, with 64 and 32 neurons in the first and second layer
respectively and with the Leaky Rectifier for activation.
In order to sample $N$ transitions with probabilities given by Equation~(\ref{Eqn:Prob}), the range [0, $\ptotal$] is divided into $N$ sub-ranges,
and a transition is uniformly sampled from each sub-range, where $\ptotal$ is the sum of priorities of all transitions in replay buffer.
As suggested by~\cite{Schaul15}, we used a sum-tree to implement the priority probability, which is similar to a binary heap. The differences are
1) leaf nodes store the priorities of transitions; and 2) internal nodes store the sum of its children.
In this way, the value of root is $\ptotal$, and the time complexity for updating and sampling is
$O(\log N_\text{tree})$, where $N_\text{tree}$ is the number of nodes in the sum-tree.
During the empirical study, we also found good settings for the other important hyper-parameters:
$\xi:=0.01$, $\beta_0:=0.6$, $\beta_1:=0.4$, $\gamma:=0.99$, $\varphi:=0.6$, $\eta^\pi:=0.001$,  $\eta^Q:=0.01$, $\tau:=0.01$ and $N=64$.

%%%%%%%%%%%%%%%%%%%%%%%%%%%%%%%%%%%%%%%%%%%%%%%%%%%%%%%%%%%%%%%%%%%%%%%%%%%%%%%%%%%%%%%%%%%%%%%
%%%%%%%%%%%%%%%%%%%%%%%%%%%%%%%%%%%%%%  Section  %%%%%%%%%%%%%%%%%%%%%%%%%%%%%%%%%%%%%%%%%%%%%%
%%%%%%%%%%%%%%%%%%%%%%%%%%%%%%%%%%%%%%%%%%%%%%%%%%%%%%%%%%%%%%%%%%%%%%%%%%%%%%%%%%%%%%%%%%%%%%%
%==========================================================
\section{Performance Evaluation}
\label{Sec:Eval}
%===========================================================

\begin{figure*}
\centering
    \begin{subfigure}[b]{0.32\textwidth}
        \includegraphics[width=\textwidth]{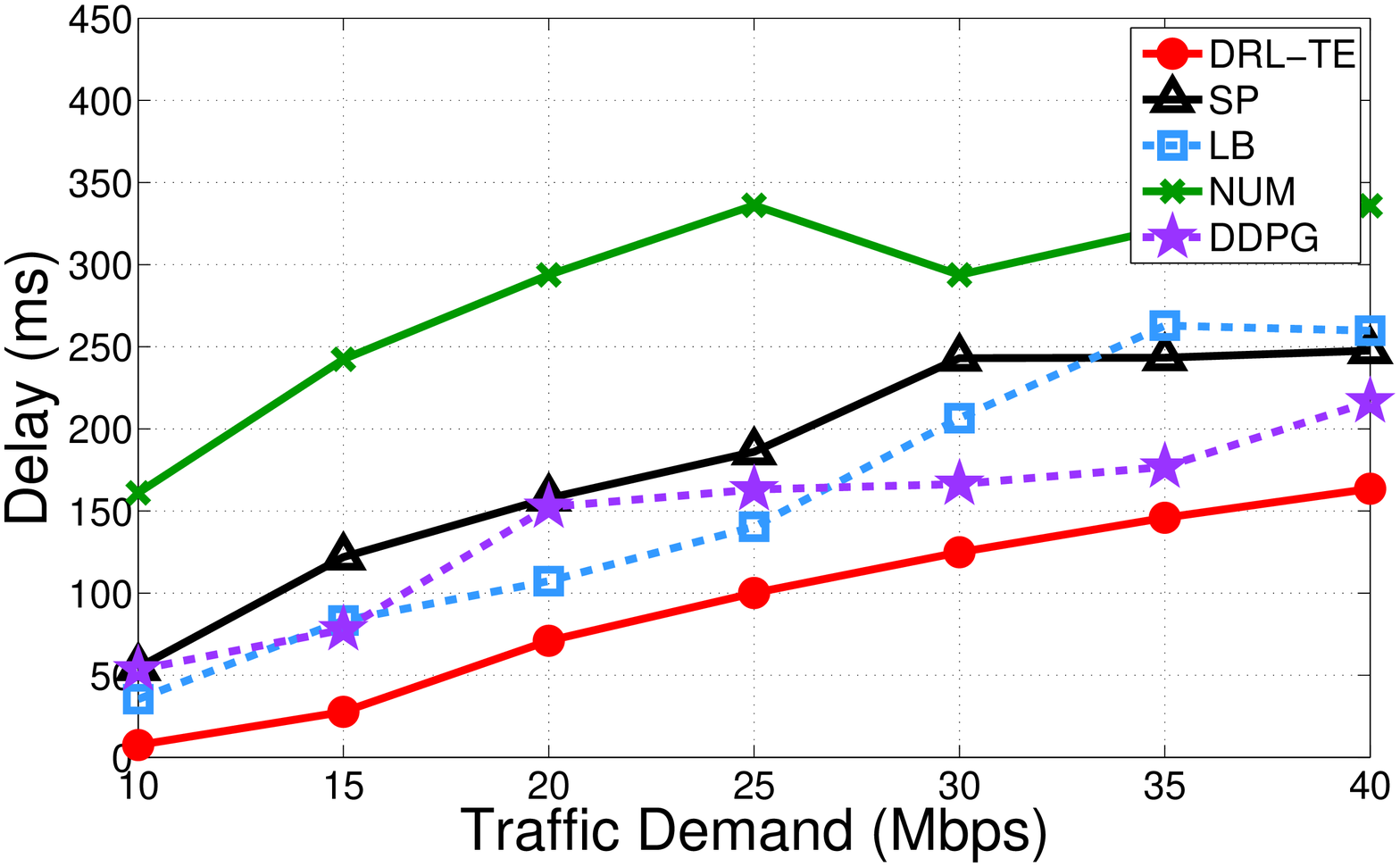}
        \caption{End-to-end delay}
        \label{fig:nsf-dly}
    \end{subfigure}
    \hfill
    \begin{subfigure}[b]{0.32\textwidth}
        \includegraphics[width=\textwidth]{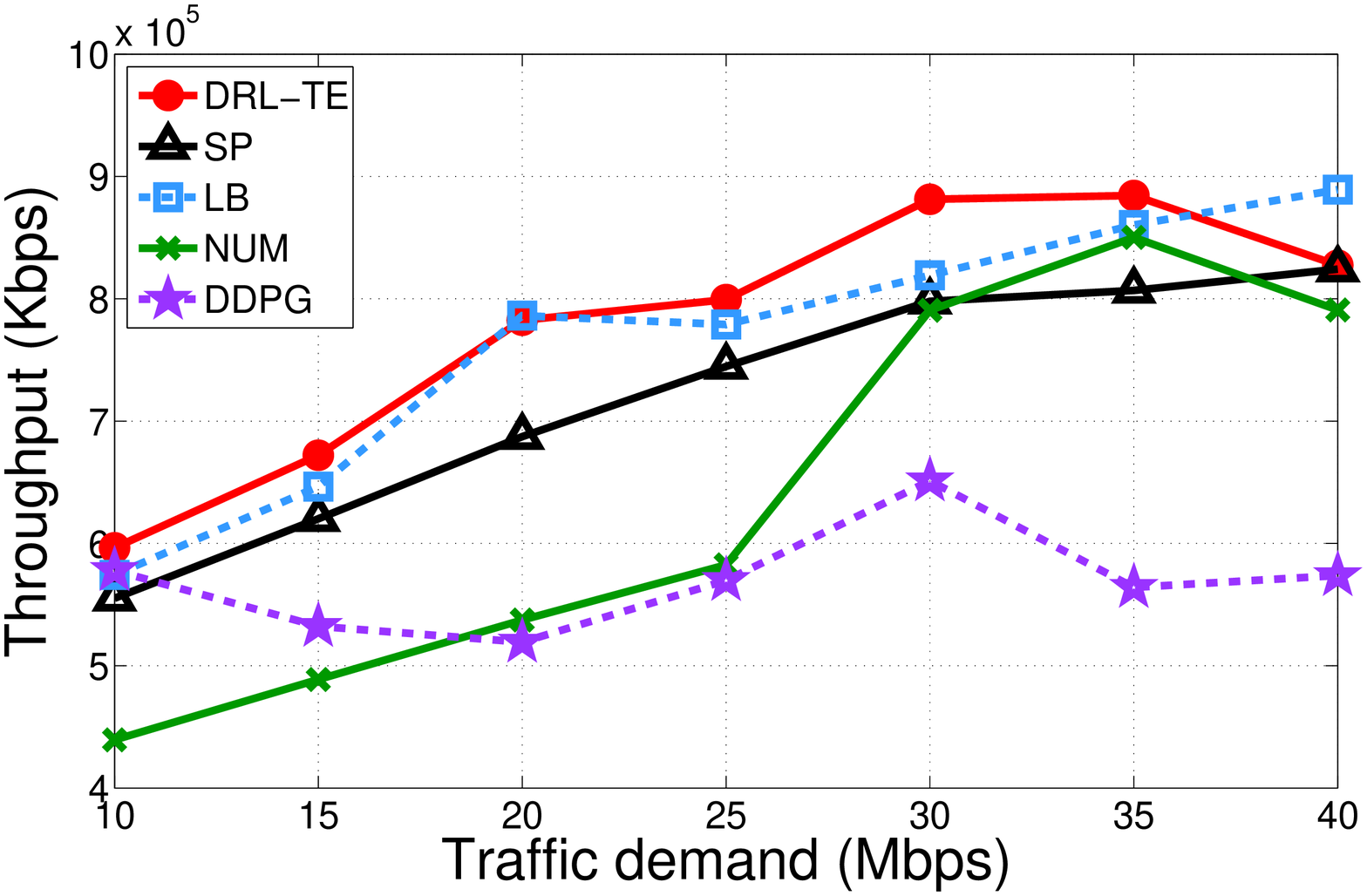}
        \caption{End-to-end throughput}
        \label{fig:nsf-thu}
    \end{subfigure}
    \hfill
    \begin{subfigure}[b]{0.32\textwidth}
        \includegraphics[width=\textwidth]{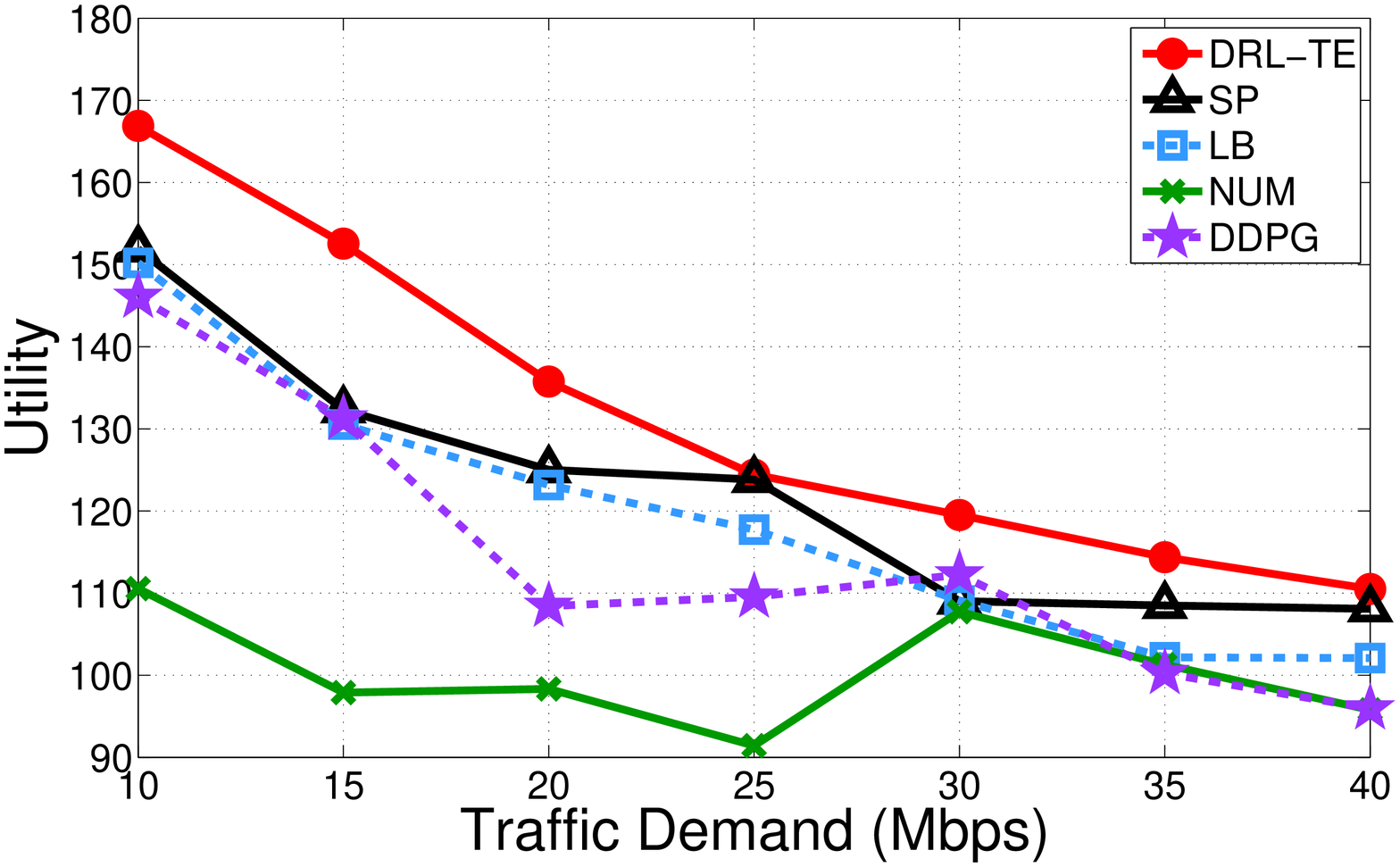}
        \caption{Network utility}
        \label{fig:nsf-utl}
    \end{subfigure}
    \caption{Performance of all the methods over the NSFNET topology}
\label{fig:nsf}
\end{figure*}

\begin{figure*}
\centering
\centering
    \begin{subfigure}[b]{0.32\textwidth}
        \includegraphics[width=\textwidth]{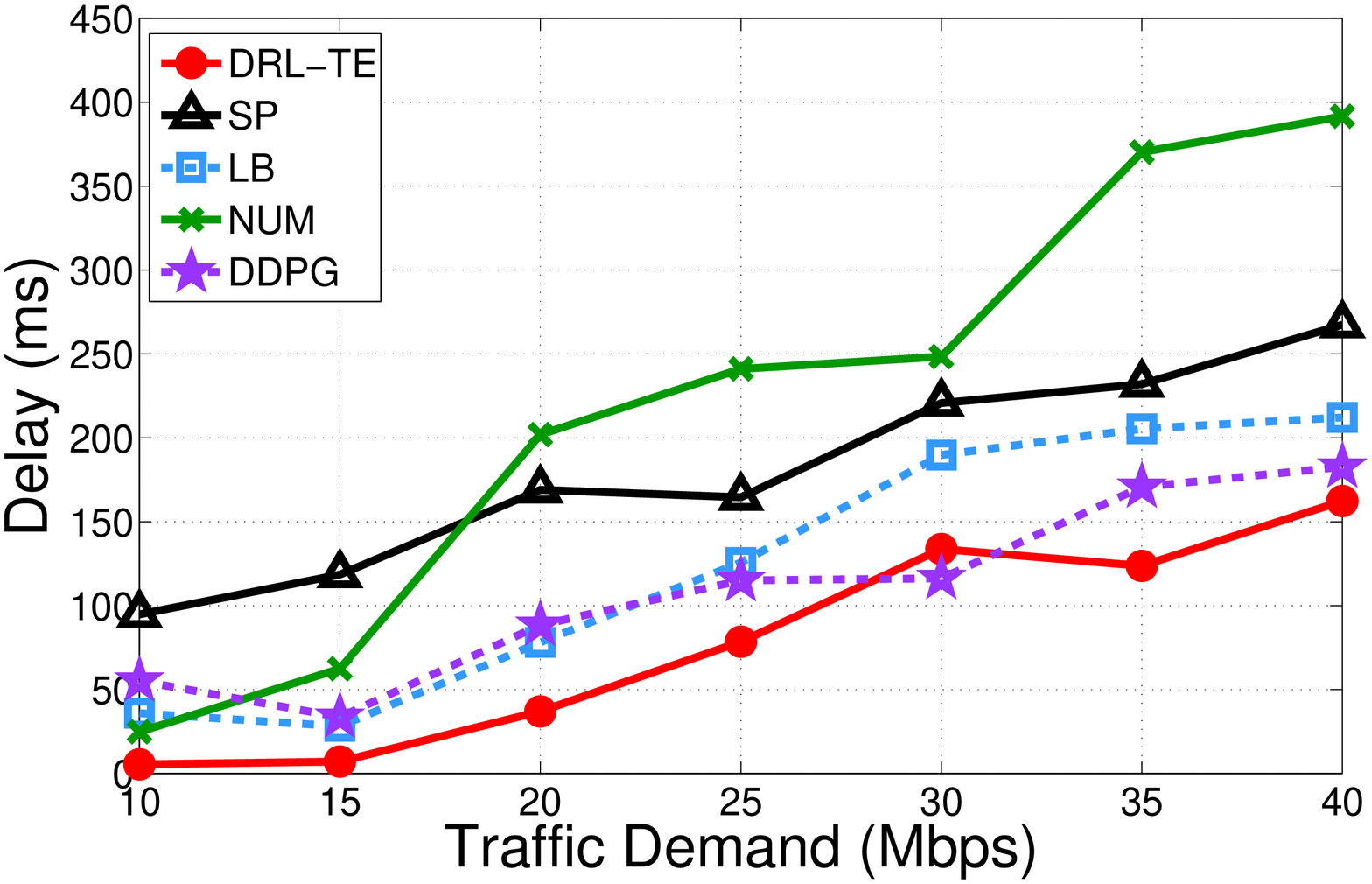}
        \caption{End-to-end delay}
        \label{fig:apa-dly}
    \end{subfigure}
    \hfill
        \begin{subfigure}[b]{0.32\textwidth}
        \includegraphics[width=\textwidth]{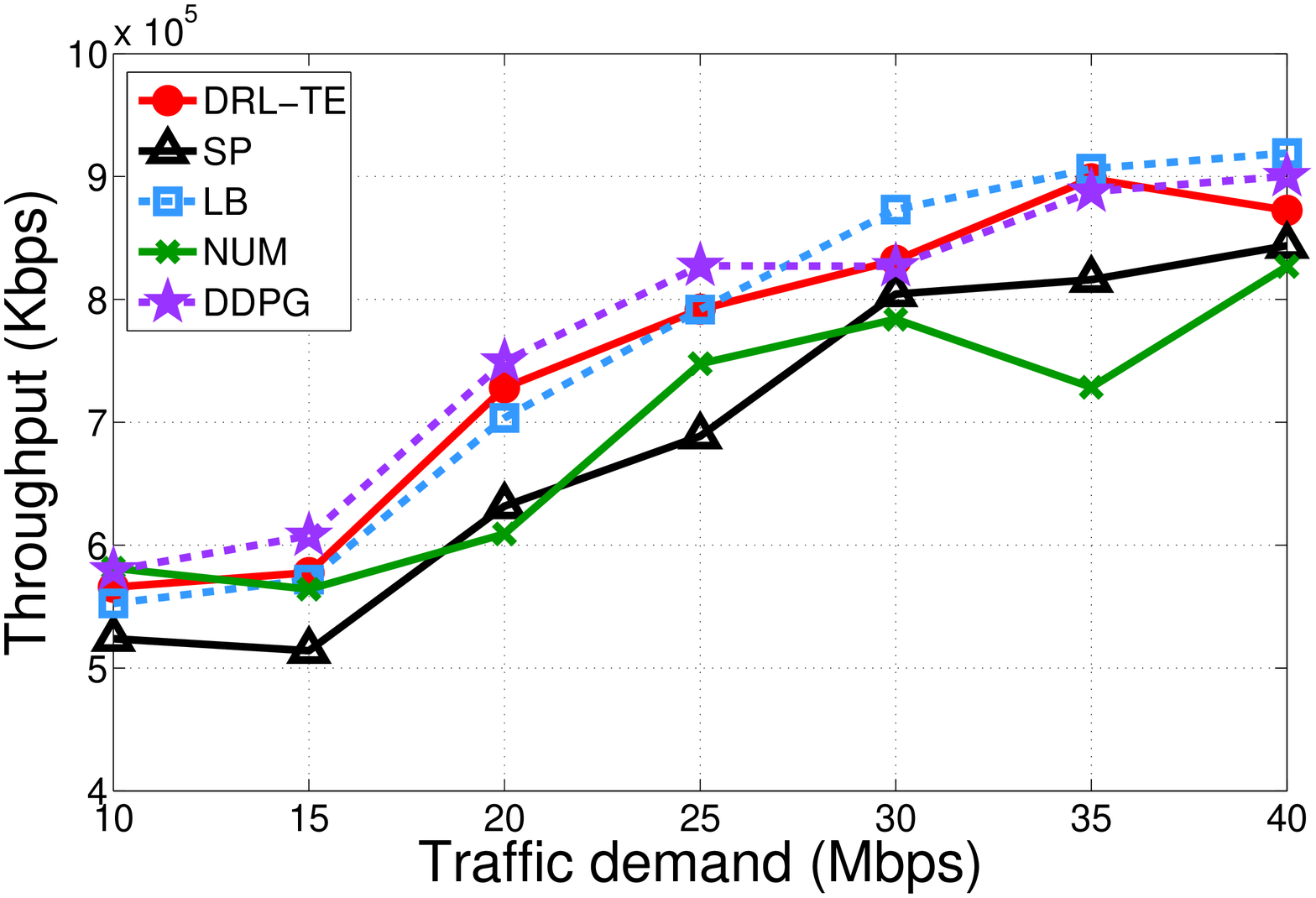}
        \caption{End-to-end throughput}
        \label{fig:apa-thu}
    \end{subfigure}
    \hfill
    \begin{subfigure}[b]{0.32\textwidth}
        \includegraphics[width=\textwidth]{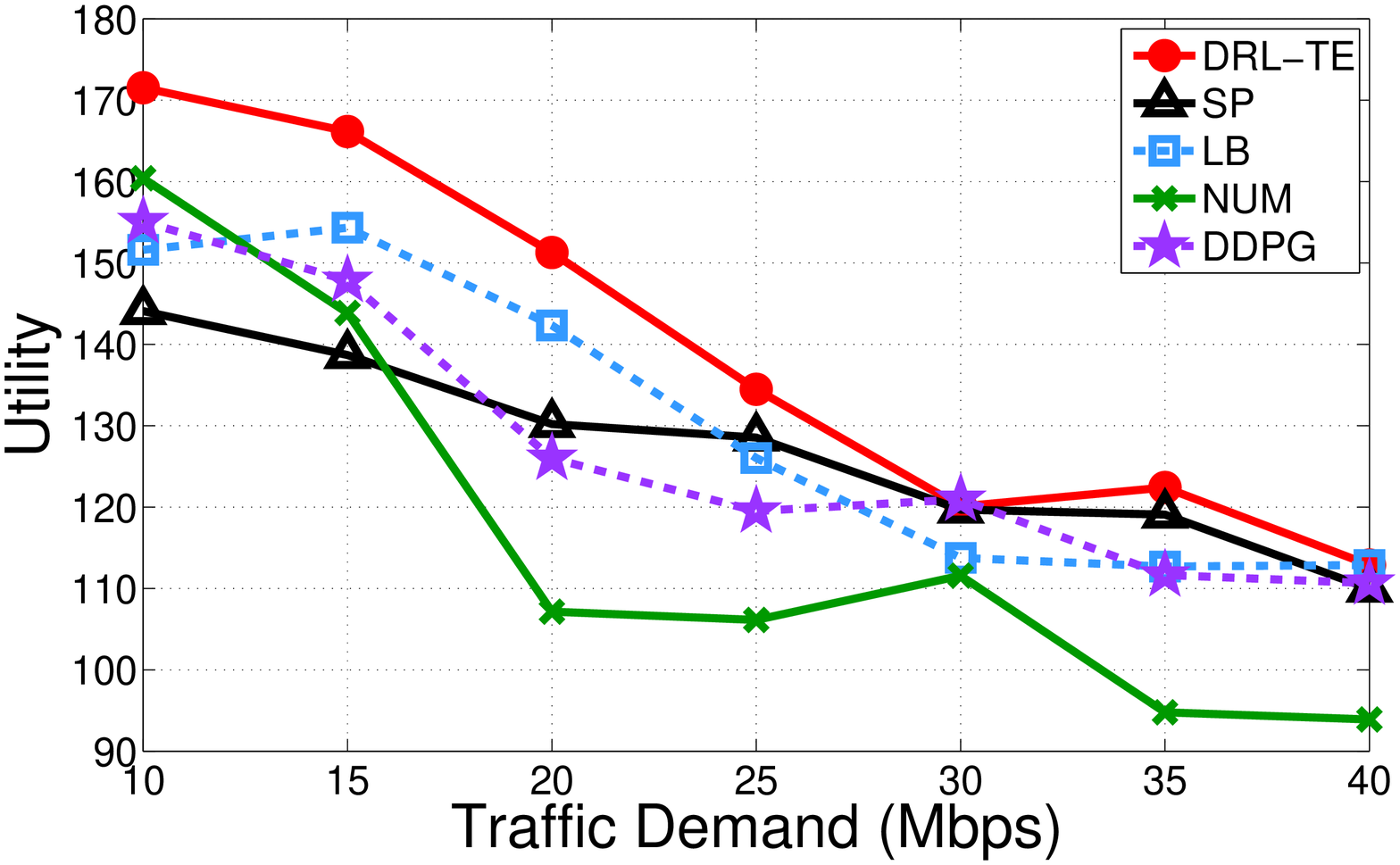}
        \caption{network utility}
        \label{fig:apa-utl}
    \end{subfigure}
    \caption{Performance of all the methods over the ARPANET topology}
\label{fig:apa}
\end{figure*}

We conducted extensive simulation to evaluate the performance of
the proposed DRL-based framework. We present and analyze the simulation
results in this section.
%
%In this section, we present simulation results to show the performance of the proposed DRL-based framework (labeled as "DRL-TE").
We implemented the proposed framework and set up the environment
in ns-3~\cite{ns3} for packet-level simulation.
%
%\Red{In the implementation, we add a extra tag to each packet to record send time,
%and at the receiver device, we analysis the tag to get the end-to-end packet delay.
%For nodes on network, we use PointToPointNetDevice to implement,
%and use PointToPointChannel to implement links on network.
%All the network parameters are set on those devices or channels,
%i.e. data rate.}
%
The DNNs included in the framework (i.e., the actor and critic networks)
were implemented using Tensorflow~\cite{TensorFlow}.
Due to the light wight of our design, we found that we could easily run and train
the proposed framework (along with the corresponding DNNs)
on a regular desktop with an Intel Quad-Core 2.6Ghz CPU with 8GB memory.

The simulation runs were performed on two well-known network topologies,
NSF Network (NSFNET~\cite{NSFNET}) and Advanced Research Projects Agency Network (ARPANET~\cite{ARPANET}).
%
%Specifically,  the NSFNET relatively small, which has a 14 nodes and 20 links,
%and ARPANET is larger, which has 20 nodes and 64 links.
%
Besides, we randomly generated a network topology with 20 nodes and 80 links,
using the widely-used network topology generator, BRITE~\cite{Medina01}.
%
%The topologies are shown in Fig.~\cite{}.
%
For each network topology, we assigned $K=20$ communication sessions,
each with randomly selected source and destination nodes.
For each communication session, we selected $3$-shortest paths (in terms of hop-count)
as its candidate paths.
The capacity of each link was set to $100$Mbps.
The packet arrival at the source node of each communication session (i.e., traffic demand) follows a Poisson process 
(note that the packet arrivals at intermediate nodes may not follow a Poisson process),
with its mean value uniformly distributed within a window with a size of $20$Mbps.
In our experiments, we set the window to $[0,20]$Mbps initially,
and we increased the traffic demand by sliding the window with a step size of $5$Mbps for each run.
We set $\alpha:=1$ and $\sigma:=1$ for the utility function to balance throughput, delay and fairness, i.e.
the objective/utility function became $\sum_{k=1}^K(\log x_k - \log z_k)$.
%
%we follow the settings of previous work~\cite{xx} to set the discount factor $\gamma=0.99$.

We compared our DRL-based control framework with three widely used baseline solutions as well as DDPG~\cite{Lillicrap16}:
\begin{itemize}
  \item Shortest Path (SP): every communication session uses a shortest path to deliver all its packets.
  \item Load Balance (LB): every communication session evenly distributes its traffic load to all candidate paths.
  \item Network Utility Maximization (NUM): it obtains TE solutions by solving the convex programming problem, NUM-TE given in Section~\ref{Sec:DRL}.
  \item DDPG: For fair comparison, we replaced the DRL-TE algorithm (Algorithm~\ref{Alg:drl}) with the DDPG algorithm~\cite{Lillicrap16}, while keeping the other settings
  (such as state, action, reward and the DNNs) the same.
\end{itemize}

\begin{figure*}
\centering
    \begin{subfigure}[b]{0.32\textwidth}
        \includegraphics[width=\textwidth]{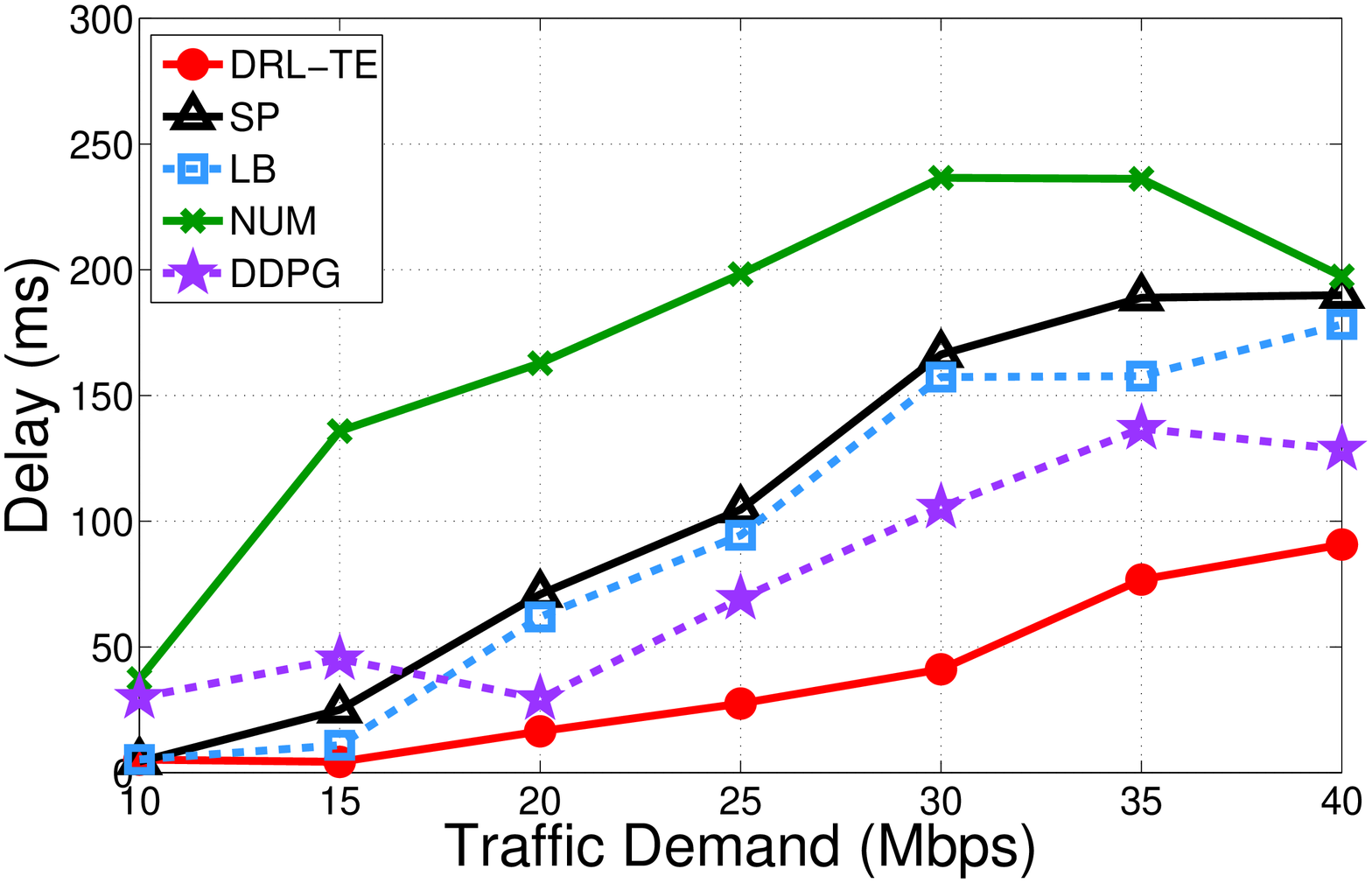}
        \caption{End-to-end delay}
        \label{fig:rad-dly}
    \end{subfigure}
    \hfill
    \begin{subfigure}[b]{0.32\textwidth}
        \includegraphics[width=\textwidth]{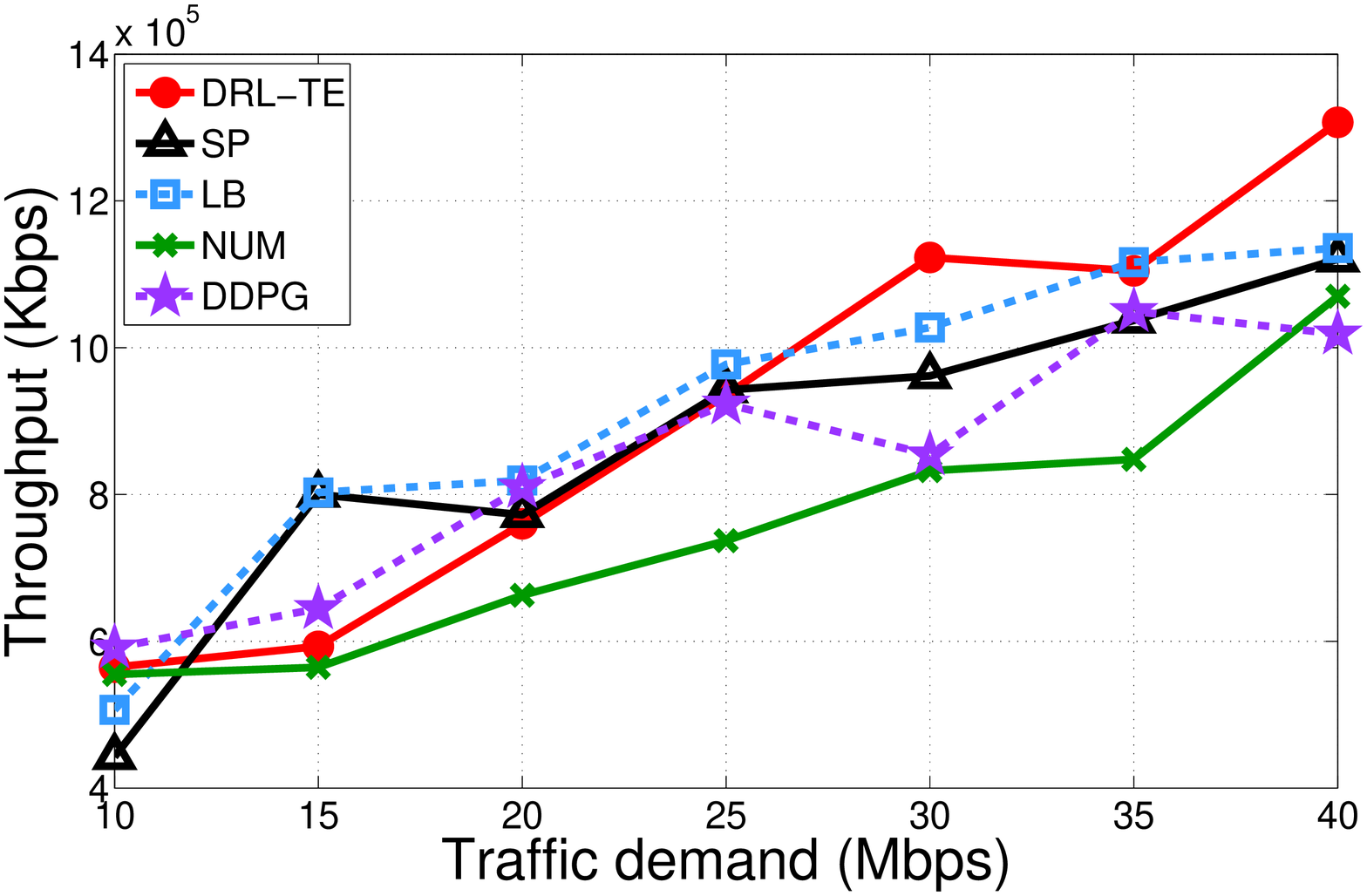}
        \caption{End-to-end throughput}
        \label{fig:rad-thu}
    \end{subfigure}
    \hfill
    \begin{subfigure}[b]{0.32\textwidth}
        \includegraphics[width=\textwidth]{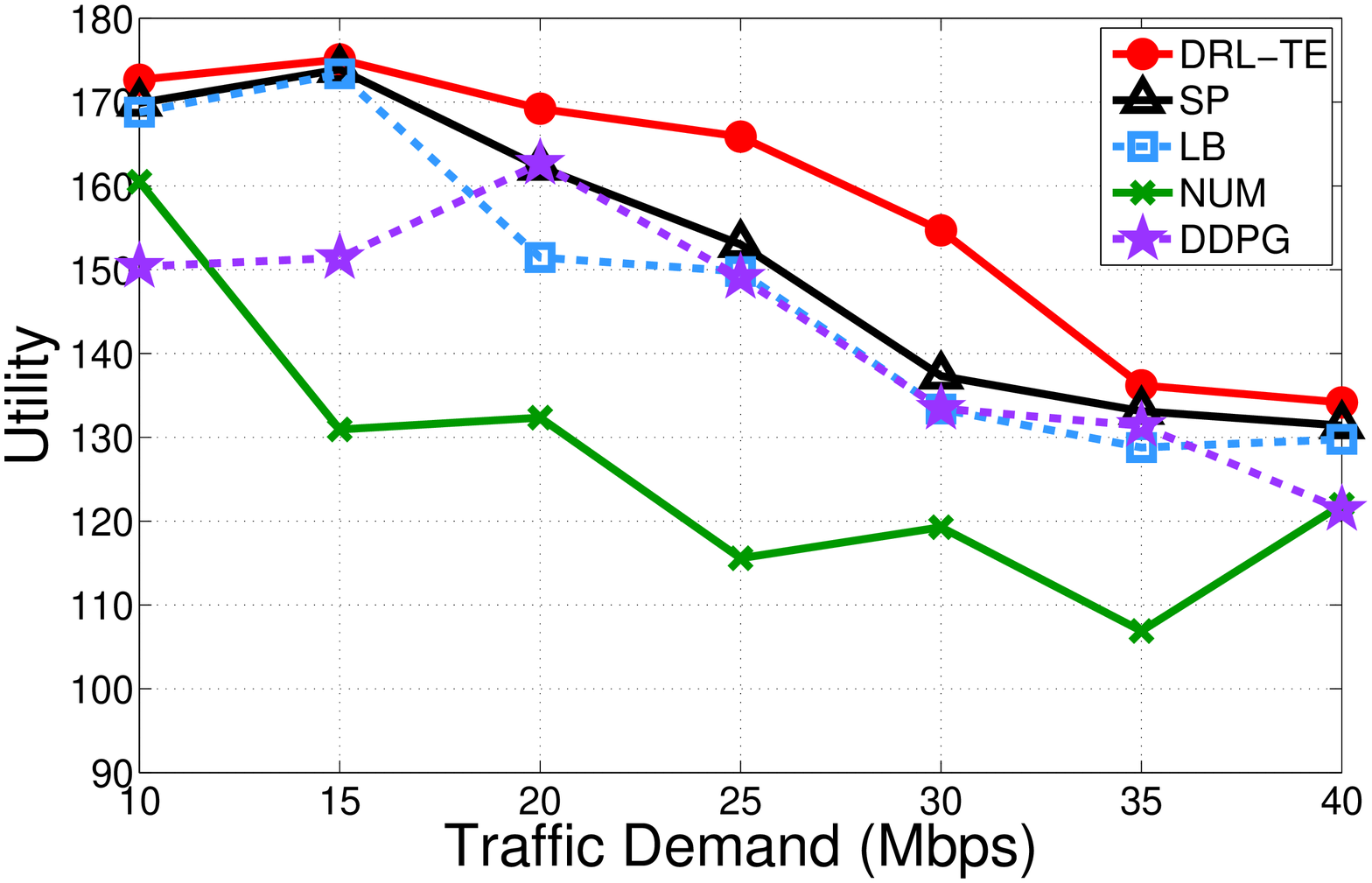}
        \caption{Network utility}
        \label{fig:rad-utl}
    \end{subfigure}
    \caption{Performance of all the methods over the random topology}
     \label{fig:rad}
   \end{figure*}

\begin{figure*}
\centering
    \begin{subfigure}[b]{0.32\textwidth}
        \includegraphics[width=\textwidth]{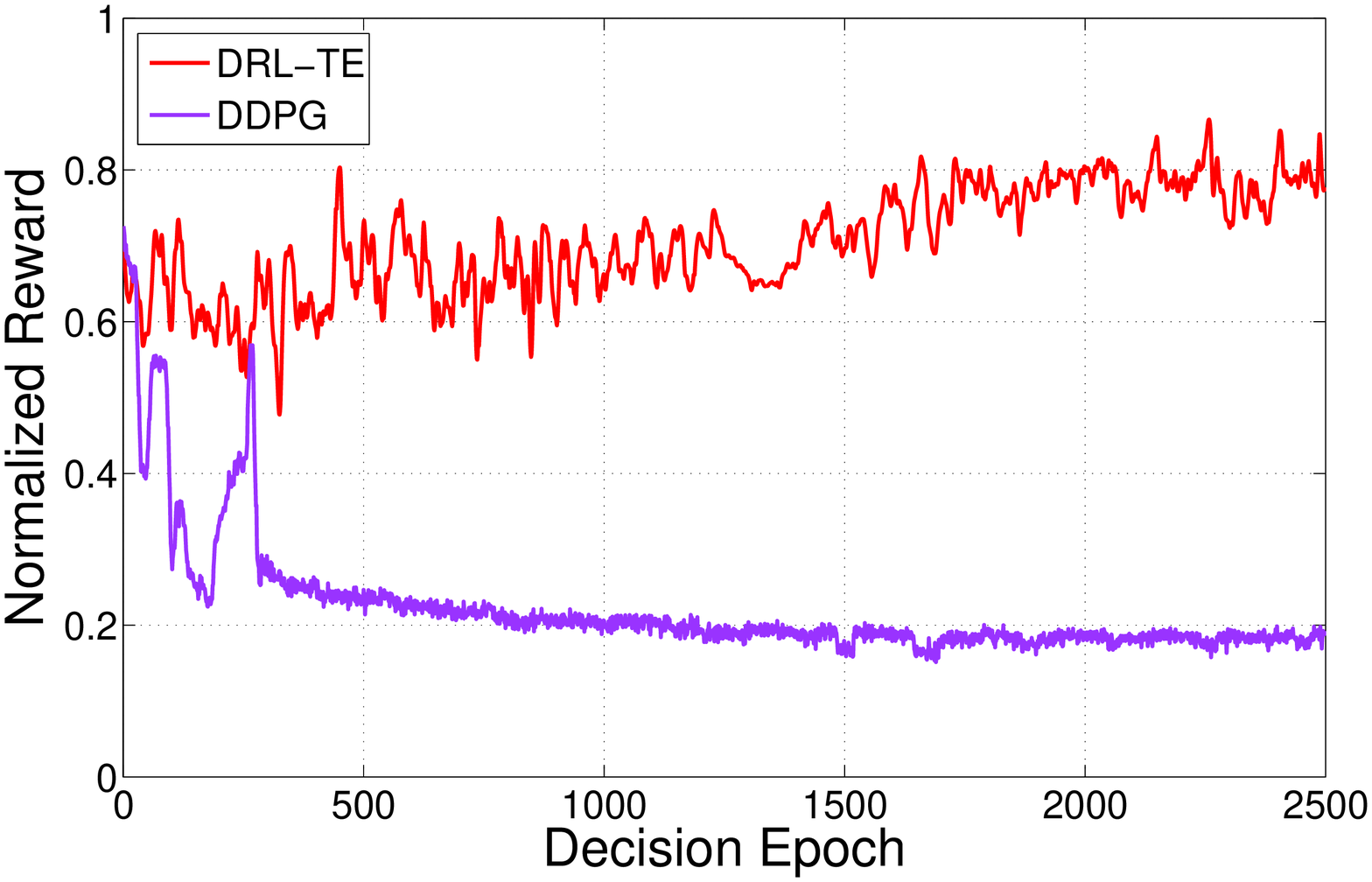}
        \caption{NSFNET topology}
        \label{fig:reward-nsf}
    \end{subfigure}
    \hfill
    \begin{subfigure}[b]{0.32\textwidth}
        \includegraphics[width=\textwidth]{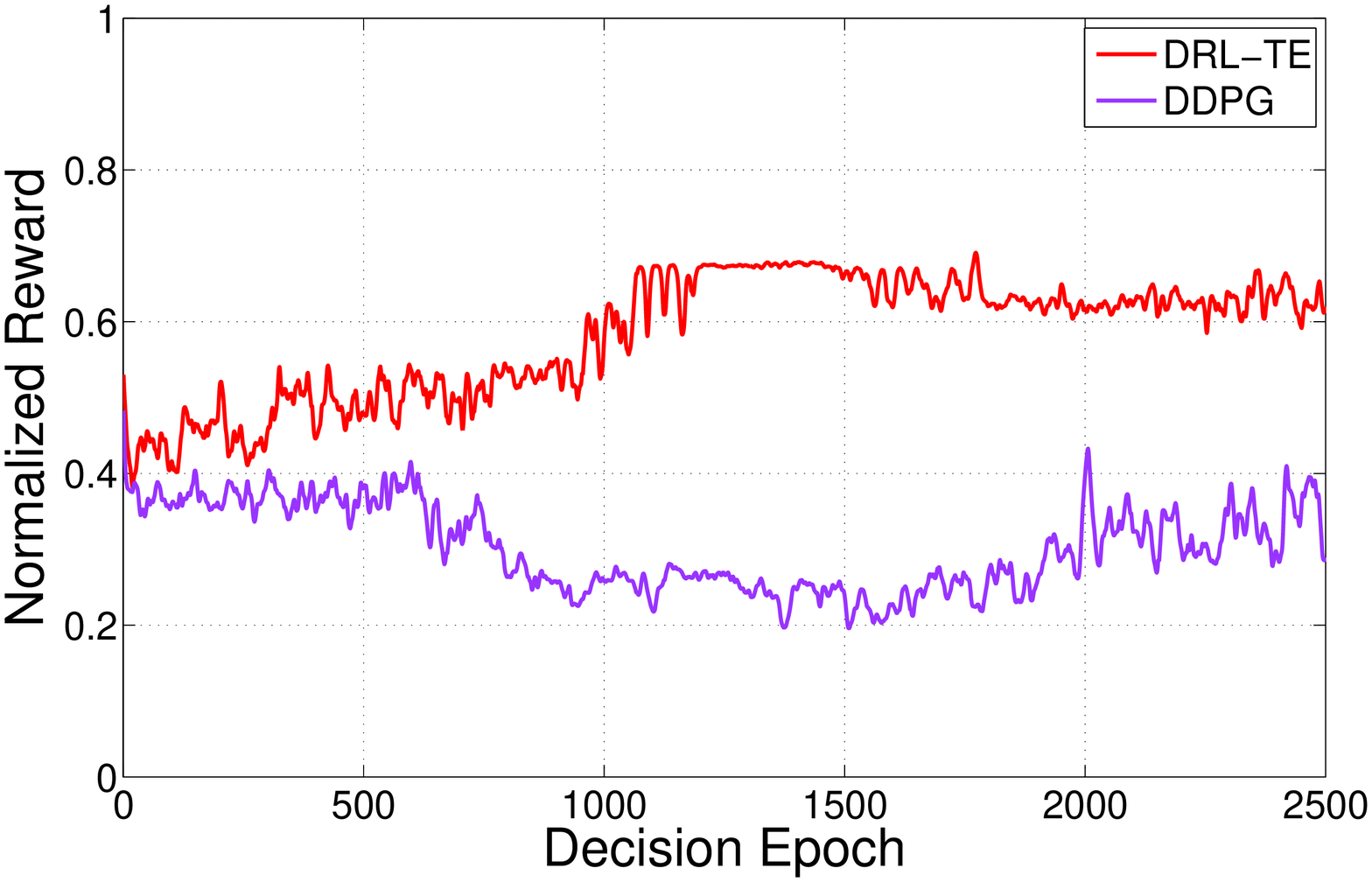}
        \caption{ARPANET topology}
        \label{fig:reward-apa}
    \end{subfigure}
    \hfill
    \begin{subfigure}[b]{0.32\textwidth}
        \includegraphics[width=\textwidth]{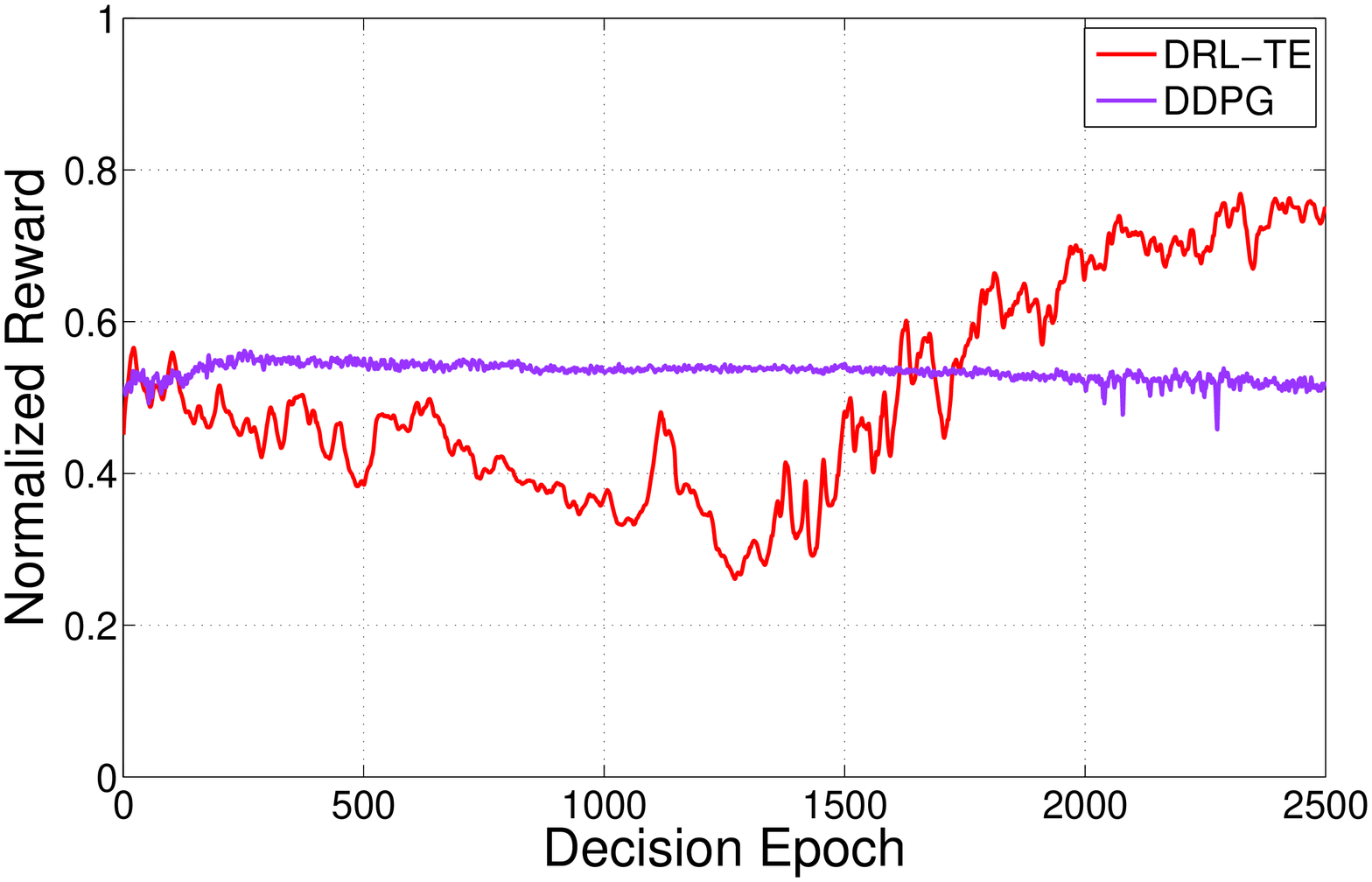}
        \caption{Random topology}
        \label{fig:reward-rad}
    \end{subfigure}
    \caption{Reward over the three network topologies during online learning}
\label{fig:reward}
\end{figure*}

We used the total end-to-end throughput, the end-to-end average packet delay, and the network (i.e., total) utility value as the performance metrics for comparisons.
%
%
%\Red{
%The throuput is caculated by the total number of bytes received divided by the time of one decision epoch.
%The delay is calculated by the average time difference between sending time and reciving time of packets, within in one decision epoch.
%}
%
%\Red{In our experiment, the decision epoch is set as $3s$ simulation time, and after $2,500$ decision epochs, the DRL agent will converge to a relative stable state.}
%
We present the corresponding simulation results in Figs.~\ref{fig:nsf}-\ref{fig:rad}, each of which corresponds to a network topology.
Note that the numbers on the x-axis are the central values of the corresponding traffic demand windows (mentioned above).
%
%As mentioned above, we performed experiments on three topologies using a sliding window traffic demand. Each point in figures corresponds to a sliding window, 20 communication sessions uniformly sample traffic demand within %that window range.
%For fair comparison, we took the final converged DRL-based solution to calculate throughput, delay and utility function, because of the learning fluctuation at the beginning.
%
In addition, we show the performance of two DRL methods (DDPG and DRL-TE) over the three network topologies during the online learning procedure in terms of the reward.
For illustration and comparison purposes, we normalized and smoothed the reward values using a commonly-used method $(r-r_{\min})/(r_{\max}-r_{\min})$ (where $r$ is the actual reward,
$r_{\min}$ and $r_{\max}$ are the minimum and maximum rewards during online learning respectively) and
the well-known forward-backward filtering algorithm~\cite{Gustafsson96} respectively. We present the corresponding simulation results
in Fig.~\ref{fig:reward}. Note that for these results, the corresponding traffic demand was generated using window $[10,30]$Mbps.
We can make the following observations from these results.

1) From Figs.~\ref{fig:nsf-dly}, \ref{fig:apa-dly} and \ref{fig:rad-dly}, we can see that compared to all the four baseline methods, DRL-TE significantly reduces end-to-end delay on all the three topologies. For example, on the NSF topology, when the traffic load is medium (i.e., traffic demand window is $[10,30]$Mbps), DRL-TE significantly reduces the end-to-end delay by $51.6\%$, $28.6\%$, $74.6\%$ and $50.0\%$ respectively, compared to SP, LB, NUM and DDPG. Overall, DRL-TE achieves an average reduction of $55.4\%$, $47.1\%$, $70.5\%$ and $44.2\%$ respectively. Compared to throughput, end-to-end delay is harder to deal with since as discussed above, it lacks accurate mathematical models that can well capture its characteristics and runtime dynamics. It is not surprising to see NUM leads to fairly poor performance since it fails to explicitly address end-to-end delay and its design is based on the assumption that network state is fairly stable or slowly changes, which may not be true; while simple solutions such as SP and LB offers expected performance since intuitively, the shortest paths and load balancing (which can avoid congestions) can help reduce delay. DRL-TE unarguably delivers superior performance with regards to end-to-end delay because it keeps learning runtime dynamics and making wise decisions to move to the optimal with the help of DNNs.

2) Even though the objective (reward function) of DRL-TE is not to simply maximize end-to-end throughput, it still delivers satisfying performance,
as shown in Figs.~\ref{fig:nsf-thu}, \ref{fig:apa-thu} and \ref{fig:rad-thu}. Compared to all the other methods, DRL-TE leads to consistently
higher throughput on the NSFNST topology. On both the ARPANET and random topologies, the throughput values given by DRL-TE are comparable to those given by LB (load balancing is supposed to yield high throughout),
but still higher than those offered by SP and NUM.

3) As expected, we can see from Figs.~\ref{fig:nsf-utl}, \ref{fig:apa-utl} and \ref{fig:rad-utl} that DRL-TE outperforms all the other methods in terms of the total utility because its reward function is set to maximizing it. On average, DRL-TE outperforms SP, LB, NUM and DDPG by $7.7\%$, $9.1\%$, $26.4\%$ and $12.6\%$ respectively. 
%
%in terms of the total utility.
%
%We can see from Figs. x-x, DRL-TE can always find a good solution which can reach a relative higher utility on different topologies. For example, on the NSFNET, DRL-TE achieves average utility improvement of 7.35\% over SP, 10.51\% over LB, and 31.45\% over NUM.

4) From Figs.~\ref{fig:nsf}-\ref{fig:rad}, we can observe no matter which method is used and no matter which network topology is chosen,
the throughput and delay basically go up with the traffic demand; while the total utility generally go down. This is easy to understand
because the higher the traffic load, usually the higher the throughput, but the higher the delay due to longer waiting time or even congestion,
which brings down the total utility. Moreover, the throughput does not increase monotonically, when the network becomes saturated,
higher traffic demands may even lead to poorer throughput due to congestion and packet losses.
We also notice that DRL-TE is robust to changes of traffic load and network topology
since it performs consistently better than all the other methods across all the traffic demand settings and all the topologies.

5) In addition, we can also observe from Figs.~\ref{fig:nsf}-\ref{fig:rad} that DDPG does not work very well on these topologies.
For example, compared to SP and LB, it performs generally worse in terms of the total utility, even though it provides slightly better end-to-end delay.
To further explain why DRL-TE works better than DDPG, we also show how the reward value changes during online learning over the three network topologies in Fig.~\ref{fig:reward}.
Clearly, over all these network topologies, DRL-TE quickly (within just a couple of thousands of decision epoches) reaches a good solution (that gives a high reward); while DDPG
seems to be stuck at local optimal solutions with lower reward values. Particularly, on the random topology, we can only see minor improvement on the first few hundred epoches, then
it fails to find better solutions (actions) to improve the reward. These results clearly justify the effectiveness of the proposed
new techniques including TE-aware exploration and the actor-critic-based prioritized experience replay.

%
%It is worth to note that the basic state-of-the-art DRL-based solution fail to find a good solution at the most of the traffic demand scenarios, and sometimes, it even gets worse than baseline solution.
%For example, on NSFNET, when the traffic demand is relatively low (i.e. median 20Mbps), DDPG get lower utility than baseline solution SP and LB. Compared with DDPG solution, we can achieve average 15.01\% utility improvement on NSFNET, ARPANET and BRITE, respectively. In the Fig. x-x, we can see that the learning reward increasing with training steps, using the same training steps, DRL-TE can quickly get more rewards than DDPG.
%
%It is interesting to see that the NUM is almost the worst solutions compared with others, but the baseline solutions (i.e. SP and LB) is not very bad on some scenarios, for example, on ARPANET, SP can even achieve the highest throughput on the high demand range.

%%%%%%%%%%%%%%%%%%%%%%%%%%%%%%%%%%%%%%%%%%%%%%%%%%%%%%%%%%%%%%%%%%%%%%%%%%%%%%%%%%%%%%%%%%%%%%%
%%%%%%%%%%%%%%%%%%%%%%%%%%%%%%%%%%%%%%  Section  %%%%%%%%%%%%%%%%%%%%%%%%%%%%%%%%%%%%%%%%%%%%%%
%%%%%%%%%%%%%%%%%%%%%%%%%%%%%%%%%%%%%%%%%%%%%%%%%%%%%%%%%%%%%%%%%%%%%%%%%%%%%%%%%%%%%%%%%%%%%%%
%==========================================================
\section{Related Work}
\label{Sec:Related}
%===========================================================
\textbf{Traffic Engineering (TE) and Network Utility Maximization (NUM):}
TE and NUM have been well studied in the literature.
%
%Steven Optimization
In a seminal work~\cite{Low99}, Low and Lapsley proposed asynchronous distributed algorithms to solve
a flow control problem whose objective is to maximize the aggregate source utility
over their transmission rates.
In~\cite{Palomar06}, Palomar and Chiang, introduced primal, dual, indirect,
partial, and hierarchical decompositions, focusing on NUM problems and the meanings of
primal and dual decompositions in terms of network architectures.
In~\cite{Paganini05}, the authors designed a congestion control system that scales gracefully with multiple objectives, which was built on decentralized control laws at end-systems.
Xu \kETAL~\cite{Xu11} proposed a new link-state routing protocol PEFT, which splits traffic over
multiple paths with an exponential penalty on longer paths, with hop-by-hop forwarding, with the objective of
achieving optimal TE.
The authors of~\cite{Li11} proposed algorithms to solve a NUM problem in a network with
delay sensitive/insensitive traffic, which is modelled by adding explicit delay terms to the utility function measuring QoS.
Einhorn \kETAL~\cite{Einhorn08} proposed a RL-based decentralized approach for QoS routing and TE in MPLS networks.
%
%Michael Delay
%The authors of ~\cite{Neely13} designed NUM algorithm that used explicit delay information from the head-of-line packet at each user.
%
Recently, TE has been studied in the context of SDN.
For example, Jain \kETAL~\cite{Jain13} presented design and implementation of
Google's SDN-based WAN, B4, and proposed a TE algorithm
based on a bandwidth function for data transmissions among its data centers.
The authors of~\cite{Agarwal13} proposed approximation algorithms for TE problems with
with partial deployment of SDN.
NUM, TE and/or related problems have also been studied by quite a few works~\cite{Bai12,Rao11,Thulasiraman11,Zhou16}
in the context of wireless networks, which were mainly focused on
wireless-specific issues such as interference, time-varying link states, etc.
We summarize the differences from these works as follows:
1) Unlike~\cite{Paganini05,Palomar06,Xu11,Li11} guided by queueing models, we develop an expereience/data-driven model-free approach based on DRL.
2) Related works~\cite{Agarwal13,Jain13,Low99} have not explicitly addressed end-to-end delay, which, however, is one of the major concerns of this paper.
3) This paper considers a TE problem in general networks, which is mathematically different from those
problems in specific networks/scenarios~\cite{Agarwal13,Bai12,Einhorn08,Jain13,Rao11,Thulasiraman11,Zhou16}.
4) We are the first to leverage the emerging DRL for TE, which has been shown to be very effective.

\textbf{Deep Reinforcement Learning (DRL):}
DRL has recently attracted extensive attention from both industry and academia.
In a pioneering work~\cite{Mnih15}, Mnih \kETAL  proposed DQN, which can learn successful policies directly
from high dimensional sensory inputs.
%
%Their work bridges the gap between high-dimensional sensory inputs and actions,
%resulting in the first artificial agent that is capable of learning to deal with a diverse array of challenging gaming tasks.
%
Particularly, they introduced two new techniques, experience replay and target network, to improve learning stability.
The authors of~\cite{Hasselt16} proposed Double Q-learning as a specific adaptation to the DQN.
%
%which is introduced in a tabular setting and can be generalized to work with a large-scale function approximation.
%
The authors of~\cite{Schaul15} proposed to use prioritized experience replay in DQN,
so as to replay important transitions more frequently, and therefore learn more efficiently.
In~\cite{Wang16}, Wang \kETAL presented a new dueling neural network architecture, which includes two separate estimators:
one for the state value function and one for the state-dependent action advantage function.
%
%Their results showed that this architecture leads to better policy evaluation in the presence
%of many similar-valued actions.
%
%The paper~\cite{Foerster16} considered a problem of multiple agents sensing and acting with the goal of maximizing their shared utility,
%based on DQN. The authors designed agents that can learn communication protocols to share information needed for accomplishing tasks.
%
The above works were focused on discrete control with a limited action space.
Research efforts have also been made to extend DRL to address continuous control.
%
%Duan \kETAL~\cite{Duan16} presented a benchmark suite of control tasks to quantify progress in the domain of continuous control.
%
Lillicrap \kETAL~\cite{Lillicrap16} proposed an actor-critic-based and model-free algorithm, DDPG, based on the deterministic
policy gradient that can operate over continuous action spaces.
%
%Other works related to this topic include~\cite{Arnold16,Gu16,Narasimhan16}.
%
Gu \kETAL~\cite{Gu16} proposed normalized advantage functions for reducing sample complexity.
%
%which enables an agent to apply Q-learning with experience replay to continuous tasks, and has been shown to substantially improve
%efficiency of simulated robotic control tasks.
%
%Arnold \kETAL~\cite{Arnold16} extended the methods proposed for continuous actions to make decisions within a large discrete action space.
%
%In~\cite{Narasimhan16}, the authors employed a DRL framework to jointly learn state representations and action policies using game rewards as feedback
%in the task of learning control policies for text-based games, where the action space contains all possible text descriptions.
%
%In~\cite{Nair15}, Nair \kETAL presented the first massively distributed architecture for DRL.
%
The authors of~\cite{Mnih16} proposed asynchronous gradient descent for optimizing learning with DNNs, and showed the successes of asynchronous the actor-critic method
on a wide variety of continuous motor control tasks.
In~\cite{Gu16Q}, the authors proposed a policy gradient method Q-Prop, which uses a Taylor expansion of the off-policy critic as a control variant.
%
%Q-Prop is both sample efficient and stable, and effectively combines the benefits of on-policy and off-policy methods.
%
We aim to answer the questions if and how the emerging DRL can be applied to solving complicated control and resource allocation problems, such as
TE, in communication networks.
Our work represents the first effort along this line. Moreover, we introduce new techniques on exploration and experience replay
to optimize the general DRL framework particularly for TE.

%================================================================
\section{Conclusions}
\label{Sec:Conclusions}
%=================================================================
In this paper, we proposed to use a novel experience-driven approach for resource allocation in communication networks,
which can learn to well control a communication network from its experience rather than an accurate mathematical model.
Specifically, we presented a novel and highly effective DRL-based control framework, DRL-TE, to solve the TE problem.
The proposed framework enables experience-driven control by jointly learning network dynamics, and make decisions
under the guidance of two DNNs, actor and critic networks.
Moreover, we proposed two new techniques, TE-aware exploration and actor-critic-based prioritized experience replay,
to optimize the general DRL framework particularly for TE.
We implemented DRL-TE in ns-3, and conducted a comprehensive simulation study to evaluate its performance
on two well-known network topologies, NSFNET and APRANET, and a random topology.
Extensive simulation results have shown that 1) compared to several widely-used baseline methods,
DRL-TE significantly reduces end-to-end delay and consistently improves the total utility, while offering
better or comparable throughput; 2) DRL-TE is robust to network changes;
and 3) DRL-TE consistently outperforms DDPG, which, however, does not offer satisfying performance.
%
%We compared our DRL-TE with several baseline solutions (i.e. SP, LB and optimization based solution NUM),
%The experiment results shown that our proposed DRL-TE can achieve higher utility and reduce delay compared with baseline solutions.
%At the same time, we showed that the direct application of DDPG does not works well.
%

%%%%%%%%%%%%%%%%%%%%%%%%%%%%%%%%%%%%%%%%%%%%%%%%%%%%%%%%%%%%%%%%%%%%%%%%%%%%%%%%%%%%%%%%%%%%%%%
%%%%%%%%%%%%%%%%%%%%%%%%%%%%%%%%%%%%%%  Section  %%%%%%%%%%%%%%%%%%%%%%%%%%%%%%%%%%%%%%%%%%%%%%
%%%%%%%%%%%%%%%%%%%%%%%%%%%%%%%%%%%%%%%%%%%%%%%%%%%%%%%%%%%%%%%%%%%%%%%%%%%%%%%%%%%%%%%%%%%%%%%
%=============================================================================

%========================================================================


\begin{thebibliography}{00}
%========================================================================
%====================================================================
\bibitem{ARPANET}
ARPANET, \emph{https://en.wikipedia.org/wiki/ARPANET}

\bibitem{Agarwal13}
S. Agarwal, M. Kodialam and TV. Lakshman,
Traffic engineering in software defined networks,
\emph{IEEE INFOCOM'2013}, pp.~2211--2219.
%
\bibitem{Arnold16}
G. D. Arnold, \kETAL,
%R. Evans, H. v. Hasselt, P. Sunehag, T. Lillicrap, J. Hunt, T. Mann, T. Weber, T. Degris and B. Coppin,
Deep reinforcement learning in large discrete action spaces,
\emph{arXiv:~1512.07679}, 2016.
%
%\bibitem{Duan16}
%Y. Duan, X. Chen, R. Houthooft, J. Schulman and P. Abbeel,
%Benchmarking deep reinforcement learning for continuous control,
%\emph{Proceedings of ICML'2016}.
%
\bibitem{Bai12}
S. Bai, W. Zhang, G. Xue, J. Tang and C. Wang,
DEAR: Delay-bounded energy-constrained adaptive routing in wireless sensor networks,
\emph{IEEE INFOCOM'2012}, pp.~1593--1601.
%
\bibitem{Einhorn08}
E. Einhorn and A. Mitschele-Thiel, RLTE: reinforcement learning for traffic-engineering,
\emph{IFIP AIMS'2008}, pp.~120--133.
%
\bibitem{Foerster16}
J. N. Foerster, Y. M. Assael, N. d. Freitas and S. Whiteson,
Learning to communicate with deep multi-agent reinforcement learning,
\emph{NIPS'2016}, pp.~2137--2145.
%
\bibitem{Goodfellow16}
I. Goodfellow, Y. Bengio and A. Courville, Deep Learning,
MIT Press, 2016, \emph{http://www.deeplearningbook.org}.
%
\bibitem{Gu16}
S. Gu, T. Lillicrap, I. Sutskever and S. Levine,
Continuous deep Q-Learning with model-based acceleration,
\emph{ICML'2016}, pp.~2829--2838.
%
\bibitem{Gu16Q}
S. Gu, T. Lillicrap, Z. Ghahramani, R. Turner and S. Levine,
Q-prop: Sample-efficient policy gradient with an off-policy critic,
\emph{arXiv:~1611.02247}, 2016.
%
\bibitem{Gurobi}
Gurobi Optimizer,
\emph{http://www.gurobi.com/}
%
\bibitem{Gustafsson96}
F. Gustafsson,
Determining the initial states in forward-backward filtering,
\emph{IEEE Transactions on Signal Processing}, Vol.~44, No.~4, 1996, pp.~988--992.
%
\bibitem{Hasselt16}
H. v. Hasselt, A. Guez, and D. Silver,
Deep reinforcement learning with double Q-learning,
\emph{AAAI'2016}, pp.~2094--2100.
%
\bibitem{Jain13}
S. Jain, \kETAL,
%A. Kumar, S. Mandal, J. Ong, L. Poutievski, A. Singh, S. Venkata, J. Wanderer, J. Zhou, M. Zhu, J. Zolla, U. Hölzle, S. Stuart and A. Vahdat
B4: Experience with a globally-deployed software defined WAN,
\emph{ACM SIGCOMM'2013}, pp.~3--14.
%
%\bibitem{Kleinrock76}
%L. Kleinrock,
%Queueing systems, volume 2: Computer applications,
%\emph{wiley New York}, Vol.~66, 1976.
%
\bibitem{Konda00}
V. Konda and J. Tsitsiklis, Actor-critic algorithms,
\emph{NIPS'2000}. pp.~1008--1014.
%
%\bibitem{Kurkowski05}
%S. Kurkowski, T. Camp and M. Colagrosso,
%MANET simulation studies: the incredibles,
%\emph{Proceedings of SIGMOBILE'2005}, pp.~50--61.
%%
%\bibitem{Lantz10}
%B. Lantz, B. Heller and N. McKeown,
%A network in a laptop: rapid prototyping for software-defined networks,
%\emph{Proceedings of SIGCOMM'2010}, pp.~19.
%
\bibitem{Li11}
Y. Li, A. Papachristodoulou, M. Chiang and A R. Calderbank,
Congestion control and its stability in networks with delay sensitive traffic,
\emph{Computer Networks}, Vol.~55, No.~1, 2011, pp.~20--32.
%
\bibitem{Lillicrap16}
T. P. Lillicrap, J. J. Hunt, A. Pritzel, N. Heess, T. Erez, Y. Tassa, D. Silver and D. Wierstra,
Continuous control with deep reinforcement learning,
\emph{ICLR'2016}.
%
%\bibitem{Lin93}
%L. Lin,
%Reinforcement learning for robots using neural networks,
%\emph{Carnegie-Mellon Univ Pittsburgh PA School of Computer Science}, 1993.
%
\bibitem{Low99}
S. H Low and D. E. Lapsley,
Optimization flow control. I. Basic algorithm and convergence,
\emph{IEEE/ACM Transactions on networking}, Vol.~518, No.~6, 1999, pp.~861--874.
%
\bibitem{McKeown08}
N. McKeown, \kETAL, OpenFlow: enabling innovation in campus networks,
\emph{ACM SIGCOMM Computer Communication Review}, Vol.~38, No.~2, 2008
pp.~69--74.
%
\bibitem{Medina01}
A. Medina, \kETAL,
%A. Lakhina, I. Matta and J. Byers,
BRITE: An approach to universal topology generation,
\emph{IEEE MASCOTS'2011}, pp~346--353.
%
\bibitem{Mnih15}
V. Mnih, \kETAL,
%
%K. Kavukcuoglu, D. Silver, A. A. Rusu, J. Veness, M. G. Bellemare, A. Graves,
%M. Riedmiller, A. K. Fidjeland, G. Ostrovski, S. Petersen, C. Beattie, A. Sadik, I. Antonoglou, H. King, D. Kumaran, D. Wierstra, S. Legg and D. Hassabis,
Human-level control through deep reinforcement learning,
\emph{Nature}, Vol.~518, No.~7540, 2015, pp.~529--533.
%
%\bibitem{Moy97}
%J. Moy, OSPF version 2, 1997.
%%
%\bibitem{Restelli15}
%M. Restelli, Reinforcement learning - exploration vs exploitation, 2015,
%\emph{http://home.deib.polimi.it/restelli/MyWebSite/pdf/rl5.pdf}
%%
\bibitem{Mnih16}
V. Mnih, \kETAL,
Asynchronous methods for deep reinforcement learning,
\emph{ICML'2016}, pp.~1928--1937.
%
%\bibitem{Nair10}
%V. Nair, and G. E Hinton,
%Rectified linear units improve restricted boltzmann machines,
%\emph{Proceedings of ICML'2010}, pp.~807--814.
%
%\bibitem{Neely13}
%M. J Neely,
%Delay-based network utility maximization,
%\emph{IEEE/ACM Transactions on Networking}, Vol.~21, No.~1, 2013, pp.~41--54.
%
\bibitem{ns3}
ns-3, \emph{https://www.nsnam.org/}
%
\bibitem{NSFNET}
NSFNET, \\
\emph{https://en.wikipedia.org/wiki/National\_Science\_Foundation\_Network}
%
\bibitem{OSPF}
Open Shortest Path First (OSPF), \\
\emph{https://en.wikipedia.org/wiki/Open\_Shortest\_Path\_First}
%
\bibitem{Paganini05}
F. Paganini, Z. Wang, J. C Doyle and S. H Low,
Congestion control for high performance, stability, and fairness in general networks,
\emph{IEEE/ACM Transactions on Networking (ToN)}, Vol.~13, No.~1, 2005, pp.~43--56.
%
\bibitem{Palomar06}
D. P. Palomar and M. Chiang, Member, A tutorial on decomposition methods for
network utility maximization, \emph{IEEE Journal on Selected Areas in Communications},
Vol.~24, No.~8, 2006, pp.~1439--1451.
%
\bibitem{Rao11}
L. Rao, X. Liu, K. Kang, W. Liu, L. Liu and Y. Chen,
Optimal joint multi-path routing and sampling rates assignment for real-time wireless sensor networks,
\emph{IEEE ICC'2011}, pp.~1--5.
%
\bibitem{Silver14}
D. Silver, G. Lever, N. Heess, T. Degris, D. Wierstra and M. Riedmiller,
Deterministic policy gradient algorithms,
\emph{ICML'2014}, pp.~387--395.
%
\bibitem{Silver16}
D. Silver, \kETAL,
%
%A. Huang, C. Maddison, A. Guez, L. Sifre, G. Van Den Driessche, J. Schrittwieser, I. Antonoglou, V. Panneershelvam, M. Lanctoc,
%S. Dieleman, D. Grewe, J. Nham,	N. Kalchbrenner, I. Sutskever,	T. Lillicrap, M. Leach,	K. Kavukcuoglu,	T. Graepel and D. Hassabis,
%
Mastering the game of Go with deep neural networks and tree search
\emph{Nature}, Vol.~529, No.~7587, 2016, pp.~484--489.
%
\bibitem{Schaul15}
T. Schaul, J. Quan, I. Antonoglou and D. Silver,
Prioritized experience replay,
\emph{arXiv:~1511.05952}, 2015.
%
\bibitem{Srikant12}
R. Srikant,
The mathematics of Internet congestion control,
\emph{Springer Science \& Business Media}, 2012.
%
\bibitem{Sutton98}
R. Sutton and A. Barto,
Reinforcement learning: an introduction,
\emph{MIT press Cambridge}, 1998.
%
\bibitem{TensorFlow}
TensorFlow,
\emph{https://www.tensorflow.org/}
%
\bibitem{Thulasiraman11}
P. Thulasiraman, J. Chen and X. Shen,
Multipath routing and max-min fair QoS provisioning under interference constraints in wireless multihop networks,
\emph{IEEE Transactions on Parallel and Distributed systems}, Vol.~22, No.~5, 2011, pp.~716--728.
%
%\bibitem{Valiant82}
%L. G Valiant, A scheme for fast parallel communication,
%\emph{SIAM journal on computing}, Vol.~11, No.~2, 1982, pp.~350--361.
%
\bibitem{Wang16}
Z. Wang, T. Schaul, M. Hessel, H. Van, M. Lanctot and N. De Freitas,
Dueling network architectures for deep reinforcement learning,
\emph{ICML'2016}, pp.~1995--2003.
%
\bibitem{Winstein13}
K. Winstein and H. Balakrishnan,
Tcp ex machina: Computer-generated congestion control,
\emph{ACM SIGCOMM'2013}, pp.~123--134.
%
\bibitem{Xu11}
D. Xu, M. Chiang and J. Rexford,
Link-state routing with hop-by-hop forwarding can achieve optimal traffic engineering,
\emph{IEEE/ACM Transactions on networking}, Vol.~19, No.~6, 2011, pp.~1717--1730.
%
\bibitem{Zhang-Shen10}
R. Zhang-Shen, Valiant Load-Balancing: building networks that can support all traffic matrices,
\emph{Algorithms for Next Generation Networks}, 2010.
%
\bibitem{Zhou16}
P. Zhou, L. Cheng and D. O. Wu, Shortest path routing in unknown environments: is the adaptive optimal strategy available?
\emph{IEEE SECON'2016}.
%

\end{thebibliography}
\end{document}